%% file: main.tex
\documentclass[article]{IEEEtran}

\IfFileExists{arXiv.tex}{\newcommand*{\arXiV}{}}{}

\ifdefined\arXiV
    \newcommand{\figures}[1]{./#1}
    
    \newcommand{\bibfolder}[1]{./#1}
    
    \newcommand{\tables}[1]{./#1}
\newcommand{\revAddB}[1]{#1}   
\newcommand{\revAddC}[1]{#1}  

\else
   \newcommand{\figures}[1]{./figures/#1}
   
   \newcommand{\bibfolder}[1]{./bib/#1}
   
    \newcommand{\tables}[1]{./tables/#1}
    \newcommand{\NoHyper}[1]{#1}
    \newcommand{\revAddB}[1]{\textbf{\textcolor{blue}{#1}}}
    \newcommand{\revAddC}[1]{\textbf{\textcolor{blue}{#1}}}
    
\fi

\usepackage[usenames,dvipsnames,svgnames,table]{xcolor}
\usepackage{pst-plot}
\usepackage{pstricks-add}

\usepackage{amsmath}
\usepackage{amssymb}
\usepackage{amsfonts}
\usepackage{cancel}
\usepackage{blindtext, graphicx}
\graphicspath{figures}
\newcommand{\figref}[1]{Figure~\ref{fig:#1}} 
\usepackage[binary-units=true]{siunitx}
\usepackage{pgfplots}
\pgfplotsset{compat=1.13}
\usepackage{pgfplotstable}
\usepackage[nomessages]{fp}
\usepackage{verbatim}
\usepackage[super]{nth}
\usepackage{soul}
\definecolor{red}{rgb}{1,0,0}
\newcommand{\new}[1]{#1}

\newcommand{\todo}[1]{\textbf{\textcolor{red}{TODO: #1}}}

\newcommand{\revComment}[1]{}

\newcommand{\revAdd}[1]{#1}
 \newcommand{\revTODO}[1]{}
 \newcommand{\revDel}[1]{}

\newcommand{\newcomment}[1]{}
\newcommand{\optprint}[1]{#1}

\usepackage{mathtools}

\DeclarePairedDelimiter{\ceil}{\lceil}{\rceil}

\usepackage[sort&compress, numbers]{natbib}

\graphicspath{{./}{./figures/}}

\ifCLASSINFOpdf
\else
\fi

\usepackage{lipsum}

\usepackage{array}
\usepackage{tabularx}
\usepackage{threeparttable}
\usepackage{booktabs}
\usepackage{colortbl}
\usepackage[binary-units=true]{siunitx}
\usepackage{bbm}
\usepackage{multirow}
\usepackage{algorithmic}
\usepackage{algorithm}
\usepackage{textcomp}

\IEEEoverridecommandlockouts
\let\footnote\thanks
\let\ACMmaketitle=\maketitle

\renewcommand{\maketitle}{\begingroup\let\footnote=\thanks \ACMmaketitle\endgroup}
\begin{document}
%
\title{YodaNN$^1$: \revAdd{An Architecture for Ultra-Low Power Binary-Weight CNN Acceleration}}

\author{\IEEEauthorblockN{Renzo Andri\IEEEauthorrefmark{1},
Lukas Cavigelli\IEEEauthorrefmark{1},
Davide Rossi\IEEEauthorrefmark{2} and
Luca Benini\IEEEauthorrefmark{1}\IEEEauthorrefmark{2}}

\IEEEauthorblockA{\IEEEauthorrefmark{1}Integrated Systems Laboratory, 
ETH Z\"urich,
Zurich, Switzerland}\\
\IEEEauthorblockA{\IEEEauthorrefmark{2}Department of Electrical, Electronic and Information Engineering, University of Bologna, Bologna, Italy}}


%


\maketitle
\stepcounter{footnote}
\footnotetext{YodaNN named after the Jedi master known from StarWars -- ``Small in size but wise and powerful'' \cite{starwars}.}
\begin{abstract}
Convolutional Neural Networks (CNNs) have revolutionized the world of computer vision over the last few years, pushing image classification beyond human accuracy. The computational effort of today's CNNs requires power-hungry parallel processors or GP-GPUs. Recent developments in CNN accelerators for system-on-chip integration have \revAddC{reduced energy consumption significantly.} Unfortunately, even these highly optimized devices are above the power envelope imposed by mobile and deeply embedded applications and face hard limitations caused by CNN weight I/O and storage. This prevents the adoption of CNNs in future ultra-low power Internet of Things end-nodes for near-sensor analytics. Recent algorithmic and theoretical advancements enable competitive classification accuracy even when limiting CNNs to binary (+1/-1) weights during training. These new findings bring major optimization opportunities in the arithmetic core by removing the need for expensive multiplications, as well as \revAddC{reducing I/O bandwidth and storage.} In this work, we present an accelerator optimized for binary-weight CNNs that achieves 1.5\,TOp/s at 1.2\,V on a core area of only 1.33\,MGE (Million Gate Equivalent) or 1.9\,mm\textsuperscript{2} and with a power dissipation of 895\,\textmu W in UMC 65\,nm technology at 0.6\,V. Our accelerator significantly outperforms the state-of-the-art in terms of energy and area efficiency achieving 61.2\,TOp/s/W\revAdd{@0.6\,V} and 1.1\,TOp/s/MGE\revAdd{@1.2\,V}, respectively.
\end{abstract}


\IEEEpeerreviewmaketitle
\bibliographystyle{IEEEtran}

%
\IEEEpeerreviewmaketitle

\input{\figures{macros.tex}}
\input{\content{1_introduction.tex}}
\input{\content{2_relatedworks.tex}}

\input{\content{3_architecture.tex}}
\input{\content{4_results.tex}}
\input{\content{5_conclusion.tex}}
\input{\content{6_futurework.tex}}
\input{\content{7_acknowledgment.tex}}



\ifCLASSOPTIONcaptionsoff
  \newpage
\fi



%

\bibliography{IEEEabrv,\bibfolder{references}}
%


\vspace{-1.5cm}
\begin{IEEEbiography}[{\includegraphics[width=1in,height=1.25in,clip,keepaspectratio]{\figures{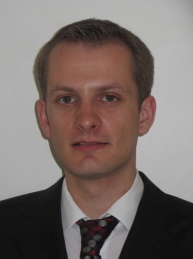}}}]{Renzo Andri} received the M.Sc. degree in electrical engineering and information technology from ETH Zurich, Zurich, Switzerland, in 2015. He is currently pursuing a Ph.D. degree at the Integrated System Laboratory, ETH Zurich. His main research interests involve the design of low-power hardware accelerators for machine learning applications including CNNs, and studying new algorithmic methods to further increase the energy-efficiency and therefore the usability of ML on energy-restricted devices.
\vspace{-1.5cm} 
\end{IEEEbiography}
\begin{IEEEbiography}[{\includegraphics[width=1in,height=1.25in,clip,keepaspectratio]{\figures{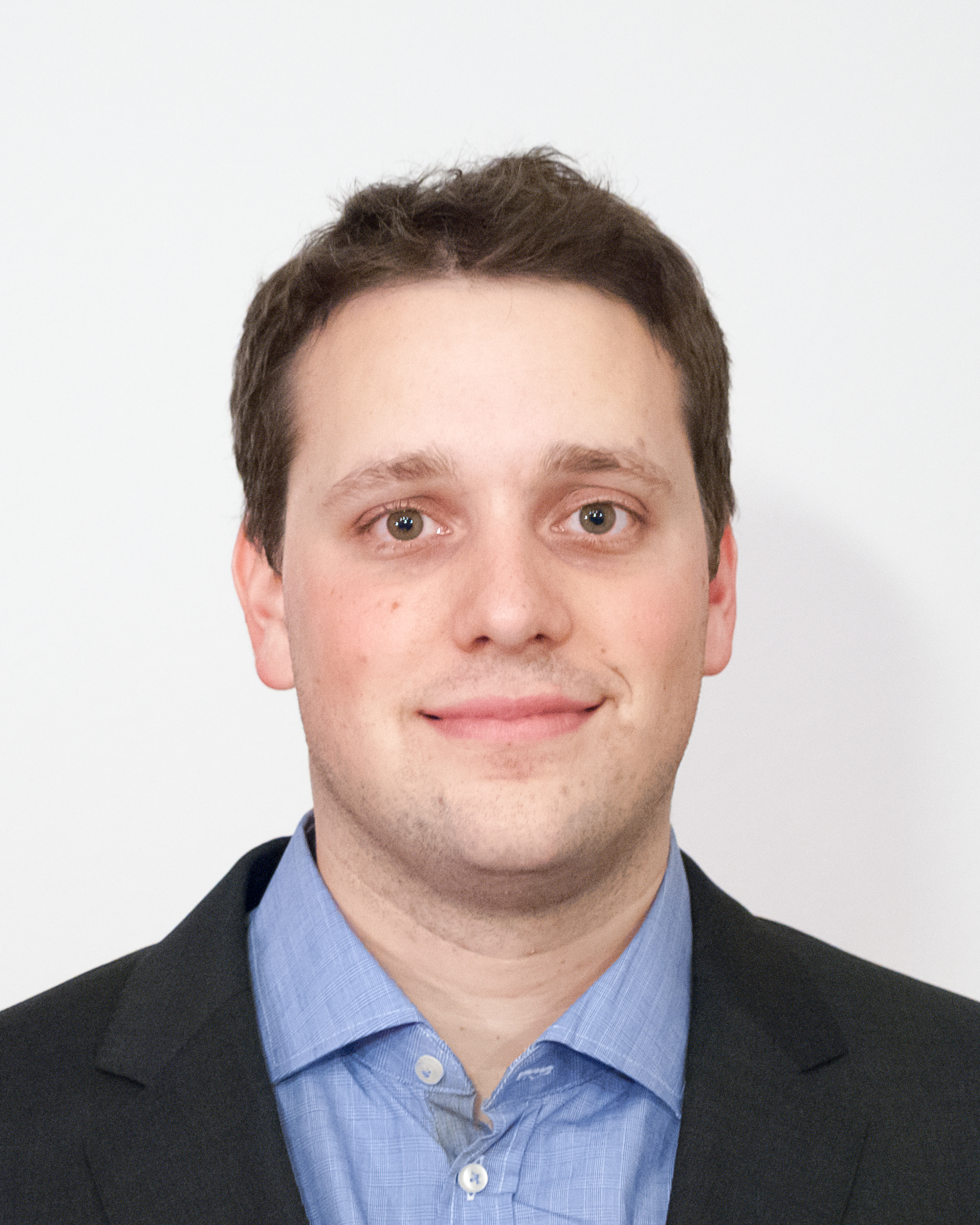}}}]{Lukas Cavigelli} received the M.Sc. degree in electrical engineering and information technology from ETH Zurich, Zurich, Switzerland, in 2014.
Since then he has been with the Integrated Systems Laboratory, ETH Zurich, pursuing a Ph.D. degree. His current research interests include deep learning, computer vision, digital signal processing, and low-power integrated circuit design.
Mr. Cavigelli received the best paper award at the 2013 IEEE VLSI-SoC Conference.
\vspace{-1.5cm}
\end{IEEEbiography}
\begin{IEEEbiography}[{\includegraphics[width=1in,height=1.25in,clip,keepaspectratio]{\figures{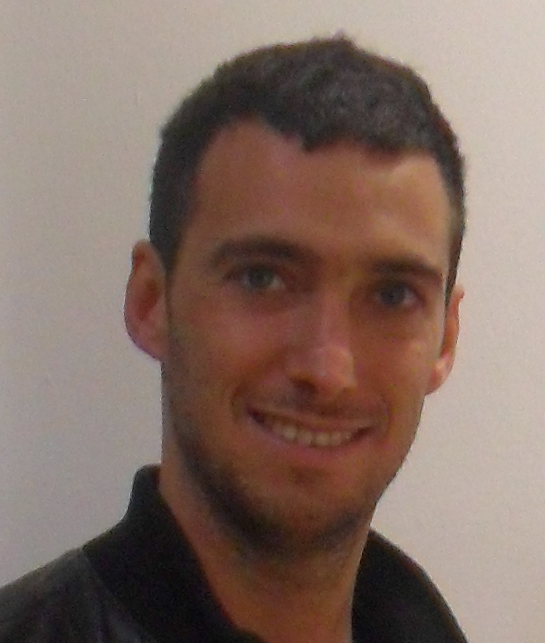}}}]{Davide Rossi} received the Ph.D. from the University of Bologna, Italy, in 2012. He has been a post doc researcher in the Department of Electrical, Electronic and Information Engineering “Guglielmo Marconi” at the University of Bologna since 2015, where he currently holds an assistant professor position. His research interests focus on energy-efficient digital architectures in the domain of heterogeneous and reconfigurable multi- and many-core systems on a chip. This includes architectures, design implementation strategies, and run-time support to address performance, energy efficiency, and reliability issues of both high end embedded platforms and ultra-low-power computing platforms targeting the IoT domain. In this fields he has published more than 30 paper in international peer-reviewed conferences and journals.
\vspace{-1.5cm}
\end{IEEEbiography}
\begin{IEEEbiography}[{\includegraphics[width=1in,height=1.2in,clip,keepaspectratio]{\figures{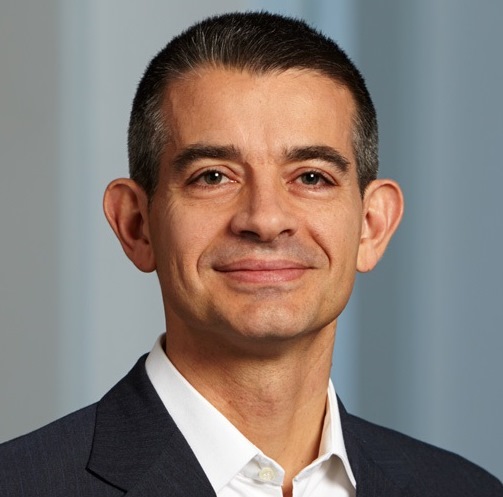}}}]{Luca Benini} is the Chair of Digital Circuits and Systems at ETH Zurich and a Full Professor at the University of Bologna. He has served as Chief Architect for the Platform2012 in STMicroelectronics, Grenoble. Dr. Benini's research interests are in energy-efficient system and multi-core SoC design.  He is also active in the area of energy-efficient smart sensors and sensor networks. 
He has published more than 700 papers in peer-reviewed international journals and conferences, four books and several book chapters. He is a Fellow of the ACM and of the IEEE and a member of the Academia Europaea. 
\vspace{-1.5cm}
\end{IEEEbiography}





\end{document}

%% file: macros.tex
\newcommand{\calc}[2]{\FPeval{\calcresult}{round(#2,#1)}\calcresult}

\newcommand{\calcsip}[2]{\FPeval{\calcresult}{round(#2,#1)}\SI{\calcresult}{\percent}}
\newcommand{\calcsi}[3]{\FPeval{\calcresult}{round(#2,#1)}\SI{\calcresult}{#3}}
\newcommand{\calcsir}[3]{\FPeval{\calcresult}{round(round(#2/(10^#1),0)*(10^#1),0)}\SI{\calcresult}{#3}}

\definecolor{color0}{RGB}{255,0,0}
\definecolor{color1}{RGB}{0,255,0}
\definecolor{color2}{RGB}{102,140,217}
\definecolor{color3}{RGB}{16,150,24}
\definecolor{color4}{RGB}{153,0,153}
\definecolor{color5}{RGB}{255,165,0}
\definecolor{cieee0}{HTML}{00629b}
\definecolor{cieee1}{RGB}{255,199,44}
\definecolor{cieee2}{RGB}{232,119,34}
\definecolor{cieee3}{RGB}{186,12,47}
\definecolor{cieee4}{RGB}{119, 37, 131}
\definecolor{cieee5}{RGB}{120, 190, 32}
\definecolor{cieee6}{RGB}{0, 132, 61}
\definecolor{cieee7}{RGB}{0,  159,  223}

\makeatletter

\tikzstyle{chart}=[
    legend label/.style={font={\scriptsize},anchor=west,align=left},
    legend box/.style={rectangle, draw, minimum size=5pt},
    axis/.style={black,semithick,->},
    axis label/.style={anchor=east,font={\tiny}},
]

\tikzstyle{bar chart}=[
    chart,
    bar width/.code={
        \pgfmathparse{##1/2}
        \global\let\bar@w\pgfmathresult
    },
    bar/.style={very thick, draw=white},
    bar label/.style={font={\bf\small},anchor=north},
    bar value/.style={font={\footnotesize}},
    bar width=.75,
]

\tikzstyle{pie chart}=[
    chart,
    slice/.style={line cap=round, line join=round, very thick,draw=white},
    pie title/.style={font={\bf}},
    slice type/.style 2 args={
        ##1/.style={fill=##2},
        values of ##1/.style={}
    }
]

\pgfdeclarelayer{background}
\pgfdeclarelayer{foreground}
\pgfsetlayers{background,main,foreground}

\newcommand{\pie}[3][]{
    \begin{scope}[#1]
    \pgfmathsetmacro{\curA}{90}
    \pgfmathsetmacro{\r}{1}
    \def\c{(0,0)}
    \foreach \v/\s in{#3}{
        \pgfmathsetmacro{\deltaA}{\v/100*360}
        \pgfmathsetmacro{\nextA}{\curA + \deltaA}
        \pgfmathsetmacro{\midA}{(\curA+\nextA)/2}

        \path[slice,\s] \c
            -- +(\curA:\r)
            arc (\curA:\nextA:\r)
            -- cycle;
        \pgfmathsetmacro{\d}{max((\deltaA * -(.5/50) + 1) , .5)}

        \begin{pgfonlayer}{foreground}
        \path \c -- node[pos=\d,pie values,values of \s]{$\v\%$} +(\midA:\r);
        \end{pgfonlayer}

        \global\let\curA\nextA
    }
    \end{scope}
}

\newcommand{\legend}[2][]{
    \begin{scope}[#1]
    \path
        \foreach \n/\s in {#2}
            {
                  ++(0,-5pt) node[\s,legend box] {} +(5pt,0) node[legend label] {\n}
            }
    ;
    \end{scope}
}

%% file: 1_introduction.tex
\section{Introduction} \label{sec:intro}

Convolutional Neural Networks (CNNs) have been achieving outstanding results in several complex tasks such as image recognition \cite{Krizhevsky2012,WuYSDS15,He2015}, face detection \cite{Taigman2014}, speech recognition \cite{HannunCCCDEPSSCN14}, text understanding \cite{Weston2014MemoryNetworks, Weston2016DialogbasedLanguageLearning} and artificial intelligence in games \cite{Mnih2015Humanlevelcontrol, Zastrow201621}. 
\revAdd{Although optimized software implementations have been largely deployed on mainstream systems \cite{Coates13}, CPUs \cite{Farabet2013} and GPUs \cite{Cavigelli2015} to deal with several state of the art CNNs, these platforms are obviously not able to fulfill the power constraints imposed by mobile and Internet of Things (IoT) end-node devices. On the other hand, sourcing out all CNN computation from IoT end-nodes to data servers is extremely challenging and power consuming, due to the large communication bandwidth required to transmit the data streams. This prompts for the need of specialized architectures to achieve higher performance at lower power within the end-nodes of the IoT.}

A few research groups exploited the customization paradigm by designing highly specialized hardware to enable CNN computation in the domain of embedded applications. Several approaches leverage FPGAs to maintain post-fabrication programmability, while providing significant boost in terms of performance and energy efficiency \cite{Ovtcharov2015}. However, FPGAs are still two orders of magnitude less energy-efficient than ASICs \cite{Cavigelli2016}. 
\revAdd{Moreover, CNNs are based on a very reduced set of computational kernels (i.e. convolution, activation, pooling), but they can be used to cover several application domains (e.g., audio, video, biosignals) by simply changing weights and network topology, relaxing the issues with non-recurring engineering which are typical in ASIC design.}

Among CNN ASIC implementations, the precision of arithmetic operands plays a crucial role in energy efficiency. Several reduced-precision implementations have been proposed recently, relying on 16-bit, 12-bit or 10-bit of accuracy for both operands and weights \cite{Cavigelli2016,Farabet:2011_NeuFlow,Conti,Du2015,Qadeer2013}, exploiting the intrinsic resiliency of CNNs to quantization and approximation \cite{Sung2015,Gysel2016}. In this work, we take a significant step forward in energy efficiency by exploiting recent research on binary-weight CNNs \cite{Courbariaux2015a, Rastegari2016}. BinaryConnect is a method which trains a deep neural network with binary weights during the forward and backward propagation, while retaining the precision of the stored weights for gradient descent optimization. This approach has the potential to bring great benefits to CNN hardware implementation by enabling the replacement of multipliers with much simpler complement operations and multiplexers, and by drastically reducing weight storage requirements. Interestingly, binary-weight networks lead to \revAddC{only }\revAdd{small} accuracy losses on several well-known CNN benchmarks \cite{Courbariaux2016,Lin2015a}.

In this paper, we introduce the first optimized hardware design implementing a flexible, energy-efficient and performance scalable convolutional accelerator supporting binary-weight CNNs. We demonstrate that this approach improves the energy efficiency of the digital core of the accelerator by $5.1\times$, and the throughput by $1.3\times$, with respect to a baseline architecture based on 12-bit MAC units \revAddC{operating at a nominal supply voltage of} \SI{1.2}{\volt}. To extend the performance scalability of the device, we implement a latch-based standard cell memory (SCM) architecture for on-chip data storage. Although SCMs are more expensive than SRAMs in terms of area, they provide better voltage scalability and energy efficiency \cite{TemanSCM2016}, extending the operating range of the device in the \revDel{high-efficiency corner}\revAdd{low-voltage region}. This further improves the energy efficiency of the engine by \calc{0}{58.56/9.61}$\times$ at \SI{0.6}{\volt}, with respect to the nominal operating voltage of \SI{1.2}{\volt}, and \revDel{leading}\revAdd{leads} to an improvement in energy efficiency by \calc{1}{58.56/5.04}$\times$ with respect to a fixed-point implementation with SRAMs at its best energy point of \SI{.8}{\volt}. To improve the flexibility of the convolutional engine we implement support for \revDel{three different kernel sizes ($3\times 3$, $5\times 5$, $7\times 7$)}\revAdd{several kernel sizes ($1\times 1$ -- $7\times 7$), and support for per-channel scaling and biasing, making it suitable for implementing a large variety of CNNs.} \revAddC{The proposed accelerator surpasses state-of-the-art CNN accelerators by 2.7$\times$ in peak performance with \SI{1.5}{\tera Op/\second} \cite{park2016}, by 10$\times$ in peak area efficiency with \SI{1.1}{TOp/s/MGE} \cite{kbrain2015} and by 32$\times$ peak energy efficiency with \SI{61.2}{\tera Op/\second/\watt} \cite{kbrain2015}.}

%% file: 2_relatedworks.tex
\section{Related Work}\label{sec:relatedwork}

Convolutional neural networks are reaching record-breaking accuracy in image recognition on small data sets like MNIST, SVHN and CIFAR-10 with accuracy rates of 99.79\%, 98.31\% and 96.53\% \cite{icml2013_wan13,2015arXiv150908985L,2014arXiv1412.6071G}. Recent CNN architectures also perform very well for large and complex data sets such as ImageNet: GoogLeNet reached \calc{2}{100-6.67}\% and ResNet \revAddC{achieved a higher recognition rate (96.43\%) than humans (94.9\%).} As the trend goes to deeper CNNs (e.g. ResNet uses from 18 up to 1001 layers, VGG OxfordNet uses 19 layers \cite{vgg2014}), both the memory and the computational complexity increases. Although CNN-based classification is not problematic when running on mainstream processors or large GPU clusters with kW-level power budgets, IoT edge-node applications have much tighter, mW-level power budgets. This "CNN power wall" led to the development of many approaches to improve CNN energy efficiency, both at the algorithmic and at the hardware level. 

\subsection{Algorithmic Approaches}\label{sec:relalgo}
Several approaches reduce the arithmetic complexity of CNNs by using fixed-point operations and minimizing the word widths. Software 
\revDel{framworks}
\revAdd{frameworks}, such as Ristretto focus on CNN quantization after training. For LeNet and Cifar-10 the additional error introduced by this quantization is less than 0.3\% and 2\%, respectively, even when the word width has been constrained to 4 bit~\cite{Gysel2016}. It was shown that state-of-the-art results can be achieved quantizing the weights and activations of each layer separately \cite{Lin2016}, while lowering precision down to 2 bit ($-1$,$0$,$+1$) and increasing the network size \cite{Sung2015}. Moons et al. \cite{leuven2016} analyzed the accuracy-energy trade-off by exploiting quantization and precision scaling. 
\revDel{Since weight distribution has peak around zero, and layer outputs contain a lot of zeros due to the often used ReLU activation function, they skip the multiplications where the weights are zero, thereby saving time and energy.}
\revAdd{Considering the sparsity in deeper layers because of the ReLU activation function, they detect multiplications with zeros and skip them, reducing run time and saving energy.} They reduce power by 30$\times$ (compared to 16-bit fixed-point) without accuracy loss, or \calc{0}{7.5*30}$\times$ with a 1\% increase in error by quantizing layers independently. 

BinaryConnect \cite{Lin2015a} proposes to binarize ($-1$, $+1$) the weights $w_{{fp}}$. During training, the weights are stored and updated in full precision, but binarized for forward and backward propagation. The following formula shows the deterministic and stochastic binarization function, where a "hard sigmoid" function $\sigma$ is used to determine the probability distribution: 
\[
    w_{b,det} = \begin{displaystyle}
  \left.
  \begin{cases}
    1, & \text{if } w_{fp}<0 \\
    -1, & \text{if } w_{fp}>0 
  \end{cases}\right.
\end{displaystyle}, \ 
w_{b,sto} = \begin{displaystyle}
  \left.
  \begin{cases}
    1, & p=\sigma(w_{fp}) \\
    -1, & p=1-\sigma  
  \end{cases}\right.
\end{displaystyle}
\]
\[
    \sigma(x)=\text{clip}\left(\frac{x+1}{2},0,1\right)=\text{max}\left(0, \text{min}\left(1,\frac{x+1}{2}\right)\right)
\]
In a follow-up work \cite{Lin2015a}, the same authors propose to quantize the inputs of the layers in the backward propagation to 3 or 4 bits, and to replace the multiplications with shift-add operations. The resulting CNN outperforms in terms of accuracy even the full-precision network. This can be attributed to the regularization effect caused by restricting the number of possible values of the weights.

Following this trend, Courbariaux et al. \cite{Courbariaux2016} and Rastegari et al. \cite{Rastegari2016} consider also the binarization of the layer inputs, such that the proposed algorithms can be implemented using only XNOR operations. In these works, two approaches are presented:
\newline i) Binary-weight-networks BWN which 
\revDel{binarize just the weights (but not the activation layers) and }
scale the output channels by the mean of the real-valued weights. With this approach they reach similar accuracy in the ImageNet data set when using AlexNet \cite{Krizhevsky2012}.
\newline ii) \revAdd{XNOR-}Networks where they also binarize the input images. This approach achieves an accuracy of 69.2\% in the Top-5 measure, compared to the 80.2\% of the setup i). Based on this work, Wu \cite{Wu2016} improved the accuracy up to 81\% using log-loss with soft-max pooling, and he was able to outperform even the accuracy results of AlexNet. However, the XNOR-based approach is not mature enough since it has only been proven on a few networks by a 
\revDel{very restricted}
\revAdd{small} research community. 
\revDel{Hence, the proposed algorithmic approach is not yet suitable for hardware implementation.} 

\revDel{The binary networks presented by Rastegari et al. use channel normalization or scaling.} 
\revAdd{Similarly to the scaling by the batch normalization,} Merolla et al. evaluated different weight projection functions where the accuracy could even be improved from 89\% to 92\% on Cifar-10 when binarizing weights and scaling every output channel by the maximum-absolute value of all contained filters \cite{Merolla2016}. In this work we focus on implementing a CNN \revAdd{inference} accelerator for neural networks supporting per-channel scaling and biasing, and implementing binary weights and fixed-point activation.\revDel{instead of the standard stochastic gradient descent algorithm,}
Exploiting this approach, the reduction of complexity is promising in terms of energy and speed, while near state-of-the-art classification accuracy can be achieved \revAdd{with appropriately trained binary networks} \cite{Courbariaux2015a, Rastegari2016}.

\subsection{CNN Acceleration Hardware}
There are several approaches to perform CNN computations on GPUs, which are able to reach a throughput up to \SI{6}{\tera Op/\second} with a power consumption of \SI{250}{\watt} \cite{Cavigelli2015,ChintalaBenchm}. On the other hand, there is a clear demand for \revAdd{low-power} CNN acceleration. For example, Google exploits in its data-centers a custom-made neural network accelerator called \textit{Tensor Processing Unit} (TPU) tailored to their TensorFlow framework. Google claims that they were able to reduce power by roughly 10$\times$ with respect to GP-GPUs \cite{TPU2016}. Specialized functional units are also beneficial for low-power programmable accelerators \revAdd{which recently} entered the market. A known example is the Movidius Myriad 2 which computes 100\,GFLOPS and needs just 500\,mW@600\,MHz \cite{Myriad2}. 
\revAdd{However, these low-power architectures are still significantly above the energy budget of IoT end-nodes.}
Therefore, several dedicated hardware architectures have been proposed to improve the energy efficiency while preserving performance, at the cost of flexibility.

Several CNN systems were presented implementing activation layer (mainly ReLU) and pooling (i.e. max pooling) \cite{kbrain2015, Jaehyeong2016, park2016}. In this work we focus on the convolution layer as this contributes most to the computational complexity \cite{Cavigelli2015}. \revDel{Convolutions have by fact}
\revAddC{Since convolutions typically rely on recent data for the majority of computations, sliding window schemes are typically used \cite{Chen2016,Conti,Jaehyeong2016,Du2015} (e.g. in case of 7$\times$7 kernels, 6$\times$7 pixels are re-used in the subsequent step).} In this work, we go even further and cache the values, such that we can reuse them when we switch from one to the next tile. In this way, only one pixel per cycle has to be loaded from the off-chip storage.


As the filter kernel sizes change from problem to problem, several approaches were proposed \revAdd{to support more than one fixed kernel size}. Zero-padding is one possibility: in Neuflow the filter kernel was fixed to $9\times 9$ and it was filled with zeros for smaller filters \cite{Pham:2012_Neuflow}. \revAdd{However, this means that for smaller filters unnecessary data has to be loaded, and that the unused hardware cannot be switched off}. Another approach 
\revDel{implements several processing elements arranged in a matrix.}
\revAdd{was presented by} Chen et al.\revAdd{, who have} proposed a\revAdd{n} accelerator containing an array of $14\times 12$ configurable  processing elements connected through a network-on-chip. The PE\revAdd{s} can be adjusted for several filter sizes. For small filter sizes, they can be used to calculate several output channels in parallel or they can be switched-off \cite{Chen2016}. Even though this approach brings flexibility, all data packets have to be labeled, such that the data can be reassembled in a later step. Hence, this system requires a lot of additional multiplexers and control logic, forming a bottleneck for energy efficiency. \revDel{In this work we propose an architecture that focuses mainly on common $3\times 3$, $5\times 5$ and $7\times 7$ kernel sizes. For different filter sizes, the filters need to be split into smaller ones and summed together off-chip.}
\revAdd{To improve the flexibility of YodaNN we propose an architecture that implements several kernel sizes ($1\times 1$, $2\times 2$, ..., $7\times 7$). Our hardware exploits a native hardware implementation for $7\times 7$, $5\times 5$, and $3\times 3$ filters, in conjunction with zero-padding to implement the other kernel sizes.}

Another approach minimizes the on-chip computational complexity exploiting the fact that due to the ReLU activation layer, zero-values appear quite often in CNNs. In this way some of the multiplications can be bypassed by means of zero-skipping \cite{Chen2016}. This approach is also exploited by Reagon et al. \cite{Reagen2016} and Albericio et al. \cite{Albericio2016}. Another approach exploits that the weights' distribution shows a clear maximum around zero.
Jaehyeong et al. proposed in their work a small 16-bit multiplier, which triggers a stall and calculation of the higher\revAdd{-order} bits only when an overflow is detected, which gives an improvement of 56\% in energy efficiency \cite{Jaehyeong2016}. \revAddC{The complexity can be reduced further by implementing } quantization scaling as described in Section~\ref{sec:relalgo}. Even though most approaches work with fixed-point operations, the number of quantization bits is still kept at 24-bit \cite{Jaehyeong2016,kbrain2015} or 16-bit \cite{Conti,Du2015, park2016,Pham:2012_Neuflow,Gokhale2014}. 

\revDel{Large bandwidth requirements are needed which affect time and energy when going off-chip.} 
\revAdd{To improve throughput and energy efficiency, }\revAdd{Hang} et al. present compressed deep neural networks, where the number of different weights are limited, and instead of saving or transmitting full precision weights, the related indices are used  \cite{Han2016}. They present a neural networks accelerator\revAdd{, called} Efficient Inference Engine (EIE), exploiting network pruning and weight sharing (deep compression). For a network with a sparsity as high as 97\%, EIE reaches an energy efficiency of \SI{5}{TOp/s/W}, and a throughput of \SI{100}{GOp/s}, which is equal to a throughput of \SI{3}{TOp/W} for the equivalent non-compressed network \cite{Han2016b}. Even though this outperforms the previous state-of-the-art by $5\times$, we can still demonstrate a $12\times $ more efficient design exploiting binary weights.\revDel{In other works compression was used, either the input data \cite{Chen2016} or the weights were compressed. E.g. } Jaehyeong et al. used PCA to reduce the dimension of the kernels. Indeed, they showed that there is a strong correlation among the kernels, which can be exploited to reduce their dimensionality without major influence on accuracy \cite{Jaehyeong2016}.\revDel{  as they approved that correlation between filters can be exploited with minor influence on accuracy .} This actually reduces the energy needed to load the chip with the filters and reduces the area to save the weights, since only a small number of bases and a reduced number of weight components need to be transmitted. On the other hand, it also increases the core power consumption, since the weights have to be reconstructed on-the-fly. With binary weights, we were able to reduce the total kernel data by $12\times$, which is similar to the 12.5$\times$ \revAddC{reported} in Jaehyeong et al. \cite{Jaehyeong2016}. On the other hand, YodaNN outperforms their architecture by $\calc{0}{61234/1420}\times$ in terms of energy efficiency thanks to its simpler internal architecture \revAddC{that do not require} on-the-fly reconstruction.
\ifdefined\FIGTABPRINT
    
\else
   \begin{figure*}
\begin{equation} \label{eq:conv}
    \mathbf{o_k} = \mathbf{C_k}+\sum_{n\in I}\underbrace{{\mathbf{i_n} \ast \mathbf{w_{k,n}}}}_{\mathbf{\tilde{o}_{k,n}}},\hspace{0.5cm}o_k(x,y)=C_k+\sum_{n\in I}\underbrace{\Bigg(\sum^{b_k-1}_{a=0}\sum^{h_k-1}_{b=0}{i_n\left(x+a,y+b\right) \cdot w_{k,n}(a,b)}\Bigg)}_{\tilde{o}_{k,n}(x,y)}
\end{equation}
\vspace{-0.8cm}
\end{figure*}
\fi
\revAdd{Some CNN accelerators have been presented exploiting analog computation: in one approach \cite{Likamwa2016}, part of the computation was  performed partially on the camera sensor chip before transmitting the data to the digital processing chip. Another mixed-signal approach [50] looked into embedding part of the CNN computation in a memristive crossbar. Efficiencies of \SI{960}{GOp/s} \cite{Likamwa2016} and \SI{380}{GOp/s/W} \cite{Shafiee2016} were achieved. YodaNN outperforms these approaches by $\calc{0}{61234/960}\times$ and $\calc{0}{61234/380}\times$ respectively, thanks to aggressive discretization and low-voltage digital logic.} 

The next step consists \revAdd{in quantizing} the weights to a binary value. However, this approach has only been \revAddC{implemented} on Nvidia GTX750 GPU leading \revAddC{to a 7$\times$ run-time reduction} \cite{Courbariaux2016}. In this work, we present the first hardware accelerator optimized for binary weights CNN, fully exploiting the benefits of the reduction in computational complexity boosting area and energy efficiency. Furthermore, the proposed design scales to deep near-threshold thanks to SCM and an optimized implementation flow, \revAddC{outperforming the state of the art by \calc{1}{1510/569}$\times$ in performance, \calc{0}{1135.3/109.6}$\times$ in area efficiency and \calc{0}{61.2/1.93}$\times$ in energy efficiency.}

%% file: 3_architecture.tex
\section{Architecture}

\begin{figure}
    \centering
    \includegraphics[width=.45\textwidth]{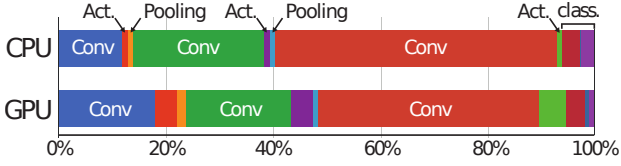}
    \caption[\revAddC{Overview of execution time in a convolution neural network for scene labeling from Cavigelli et~al. executed on CPU and GPU.}]{\revAddC{Overview of execution time in a convolution neural network for scene labeling from Cavigelli et~al. executed on CPU and GPU \cite{Cavigelli2015}.}}
    \label{fig:cnn_vs_others}
\end{figure}
\begin{figure}
    \centering
    \def\svgwidth{0.44\textwidth}
    \graphicspath{./figures/cnn_overview_detail}
    \resizebox{!}{0.29\textheight}{\resizebox{0.45\textwidth}{!}{\input{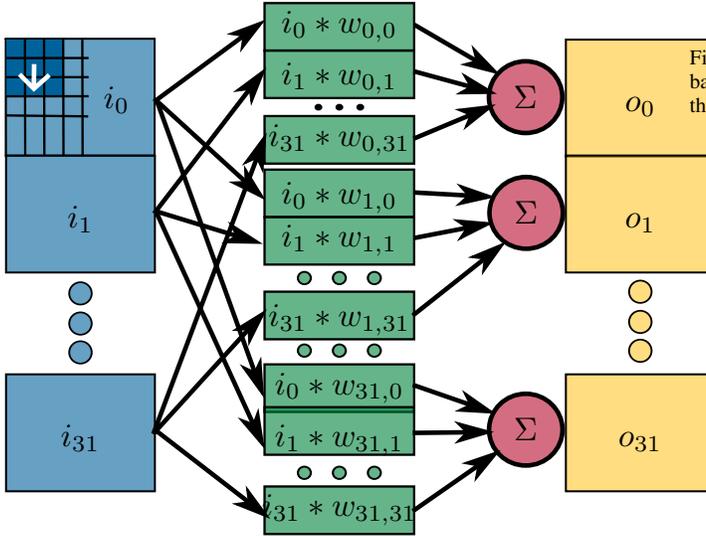}}}
    \caption{A 32$\times$32 CNN layer, with input channels $i_n$ and output channels $o_k$.}
   \label{fig:cnn}
\end{figure}

\begin{figure}
    \centering
    \def\svgwidth{0.48\textwidth}
    \input{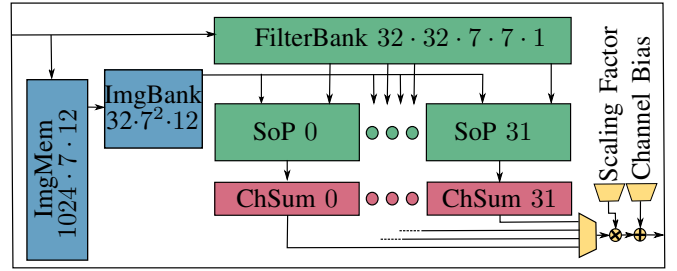}
    \caption{General overview of the system with the image memory and image bank in blue, filter bank and SoP units in green, channel summer in red and the interleaved per-channel scaling, biasing and streaming-out units in yellow.}
   \label{fig:overalloverview}
\end{figure}

A CNN consists of several \revAdd{layers}, usually they are convolution, activation, pooling or batch normalization layers. In this work, we focus on the convolution layers as they make up for the largest share of the total computation time. \revAddC{As can be seen in \figref{cnn_vs_others} from \cite{Cavigelli2015}, convolution layers make up for the largest fraction of compute time in CPU and GPU implementations. This is why we focus on convolution layers in this work.} A general convolution layer is drawn in \figref{cnn} and it is described by Equation \eqref{eq:conv}. A layer consists of $n_{in}$ input channels, $n_{out}$ output channels, and $n_{in}\cdot n_{out}$ kernels with $h_k\times b_k$ weights; we denote the matrix of filter weights as $w_{k,n}$. For each output channel $k$ every input channel $n$ is convolved with a different kernel $w_{k,n}$, resulting in the terms $\tilde{o}_{k,n}$, which are accumulated to the final output channel $o_k$.
We propose a hardware architecture able to calculate $n_{ch}\times n_{ch}$ channels in parallel. If the number of input channels $n_{in}$ is greater than $n_{ch}$, the system has to process the network $\ceil{n_{in}/n_{ch}}$ times and the results are accumulated off-chip. This adds only $\ceil{n_{in}/n_{ch}}-1$ operations per pixel. In the following, we fix, for ease of illustration, the number of output channels to $n_{ch}=32$ and the filter kernel size to $h_k=b_k=7$. The system is composed of the following units (an overview can be seen in \figref{overalloverview}):

\begin{itemize}
    \item The \textit{Filter Bank} is a shift register which contains \revAddC{the binary} filter weights $w_{k,n}$ for the output channels $k\in \mathbb{N}_{<32}$ and input channels $n\in \mathbb{N}_{<32}$ ($n_{in}\cdot n_{out}\cdot h_k^2\cdot 1\,\text{bit}=6.4\,\text{kB}$) and supports column-wise left circular shift per kernel. 
    \item The \textit{Image Memory} saves an image stripe of $b_k=7$ width and $1024$ height (\calcsi{1}{1024*7*12/1000/8}{\kilo\byte}), which can be used to cache $1024/n_{in}=1024/32=32$ rows per input channel.
    \item The \textit{Image Bank} (ImgBank) caches a spatial window of $h_k \times b_k=7\times 7$ per input channel $n$ (\calcsi{1}{32*7*7*12/1000/8}{\kilo\byte}), which are applied to the SoP units. This unit is used to reduce memory accesses, as the $h_k-1=6$ last rows can be reused when we proceed in a column-wise order through the input images. Only the lowest row has to be loaded from the image memory and the upper rows are shifted one row up.
    \item \textit{Sum-of-Product} (SoP) Units \revAddC{(32, 1 per output channel)}: \revAddC{For every output channel $k$, the }SoP unit $k$ calculates the sum terms $\tilde{o}_{k,n}$, where in each cycle the contribution of a new input channel $n$ is calculated.
    \item \textit{Channel Summer} (ChSum) Units \revAddC{(32, 1 per output channel)}: The Channel Summer $k$ accumulates the sum terms $\tilde{o}_{k,n}$ for all input channels $n$.
    \item \new{\textit{$1$ Scale-Bias Unit}: \revAdd{After all the contributions of the input channels are summed together by the channel summers, this unit starts to scale and bias the output channels in an interleaved manner and streams them out.} \revDel{In an interleaved manner the channel sums are processed in the Scale-Bias Unit. The sums are truncated to the initial fixed-point format Q2.9 (12\,bit), scaled and biased based on the current channel, and then finally streamed out.}}
    \item \revAdd{\textit{I/O Interface}: Manages the 12-bit input stream (input channels) and the two 12-bit output streams (output channels) with a protocol based on a blocking ready-valid handshaking.}
\end{itemize}

 %

\begin{figure*}
    \centering
    \def\svgwidth{1.4\textwidth}
    \resizebox{1\textwidth}{!}{\hspace{-0.85cm}\input{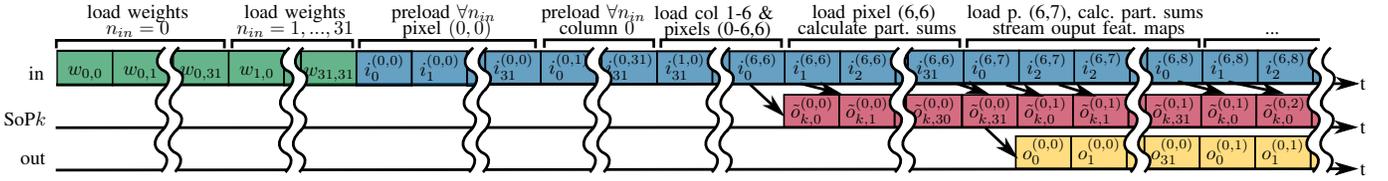}}
    \caption{Timing diagram of the operating scheme: Input Stream, SoP $k$'s operations, output stream after accumulation.}
   \label{fig:timing}
\end{figure*}

\subsection{Dataflow}\label{sec:dataflow}
\begin{algorithm}                      
\caption{Dataflow Pseudo-Code}          
\label{alg1}                           
\begin{algorithmic}[1]                    
    \REQUIRE weights $w_{k,n}$, input feature map $i_k(x,y)$
    \ENSURE $o_n = \sum_{k}{i_k\ast w_{k,n}}$
    \FORALL{$y_{block} \in \{1, .., \lceil h_{im}/h_{max}\rceil\}$}
        \FORALL{$c_{out, block} \in \{1, .., \lceil n_{out}/n_{ch}\rceil\}$}
            \FORALL{$c_{in, block} \in \{1, .., \lceil n_{in}/n_{ch}\rceil\}$}
                \STATE \textit{-- YodaNN chip block}
                \STATE Load Filters $w_{k,n}$
                \STATE Load $m$ colums, where \newline$m=\begin{cases} 
      h_k-1, & \text{if not zero-padded} \\
      \lfloor \frac{h_k-1}{2}\rfloor, & \text{if zero-padded} 
    \end{cases}$
                \STATE Load $m$ pixels of the $(m+1)^\text{th}$ column.
                \STATE
                \STATE{-- Parallel block 1}
                \FORALL{x}
                    \FORALL{y}
                        
                        \STATE $\tilde{o}(c_{out}:=\cdot, x,y) := 0$
                        \FORALL{$c_{in}$}
                            \STATE
                            \STATE \textit{\textbf{-- Single cycle block}}
                            
                            \FORALL{$c_{out}$}
                                \FORALL{(a,b) $\in$ $\{-\lfloor\frac{h_k}{2}\rceil\le a,b\le \lceil\frac{h_k}{2}\rfloor\}$}
                                    \STATE $\tilde{o}_{c_{out}}(x,y) = \tilde{o}_{c_{out}}(x,y) +$\newline
                                    $\text{\ \ \ \ \ \ } $\ \ $i_{c_{in}}(x\text{+}a, y\text{+}b)\cdot w_{c_{out}, c_{in}}(a,b)$ 
                                \ENDFOR
                            \ENDFOR
                        \ENDFOR
                        
                    \ENDFOR
                \ENDFOR
                \STATE{-- Parallel block 2}
                \FORALL{x}
                    \STATE{\textbf{wait until} $\tilde{o}_0(x,0)$ is computed}
                    \FORALL{y}
                            
                            \FORALL{$c_{out}$}
                                \STATE \textit{\textbf{-- Single cycle block}}
                                \STATE $o_{c_{out}}(x,y)=\alpha_{c_{out}} \tilde{o}_{c_{out}}(x,y)+\beta_{c_{out}}$
                                \STATE \textbf{output} $o_{c_{out}}(x,y)$
                            \ENDFOR
                    \ENDFOR
                \ENDFOR
                    
            \ENDFOR
            \STATE -- Sum the input channel blocks:
            \STATE $o_{n,final} = \sum_{c_{in,blocks}}{o_{n, \cdot}}$
        \ENDFOR
    \ENDFOR
\end{algorithmic}
\end{algorithm}

The pseudo-code in Algorithm~1 gives an overview of the main steps required for the processing of convolution layers, while \figref{timing} shows a timing diagram of the parallel working units. The input and output channels need to be split into blocks smaller than $32\times 32$, while the image is split into slices of $1024/c_{in}$ height (lines 1--3). These blocks are indicated as \textit{YodaNN chip block}. \revDel{Then, the 1-bit filters are loaded into the chip (Line~5 and in green). Some columns have to be preloaded, as they are needed to calculate a valid result,} \revAddC{Depending on whether the border is zero-padded or not,} $\lfloor (h_k-1)/2\rfloor$ or $h_k-1$ columns need to be preloaded (just in case of $1\times 1$ filters no pixels need to be preloaded) (Line~6). The same number of pixels are preloaded from one subsequent column, such that a full square of $h_k^2$ pixels for each input channel is available in the image bank (Line~7). After this preloading step, the SoPs start to calculate the partial sums of all 32 output channels while the input channel is changed every cycle (lines 15--20). When the final input channel is reached, the channel summers keep the final sum for all 32 output channels of the current row and column, which are scaled and biased by the Scale-Bias Unit and the final results are streamed out in an interleaved manner (lines 27--33). In case of $n_{out}=n_{in}$ (e.g. $32\times 32$) the same number of cycles are needed to stream out the pixels for all output channels as cycles are needed to sum all input channels for the next row, which means \revAddC{that all computational units of} the chip are \revAddC{fully-utilized}.
\revAddC{Each row is processed sequentially, }then the system switches to the next column, where again the first pixels of the column are preloaded. The filters are circularly right shifted to \revAdd{be aligned to the correct columns.}
\revAdd{Then, the next column of all output channels are calculated. } \revAdd{This procedure is repeated until the whole image and blocks of input and output channels have been processed.} \revAdd{Finally, the partial sums for each output channels need to be summed together for every block of input channels. (Line~37).}

\begin{figure}
    \centering
    \includegraphics[width=0.45\textwidth]{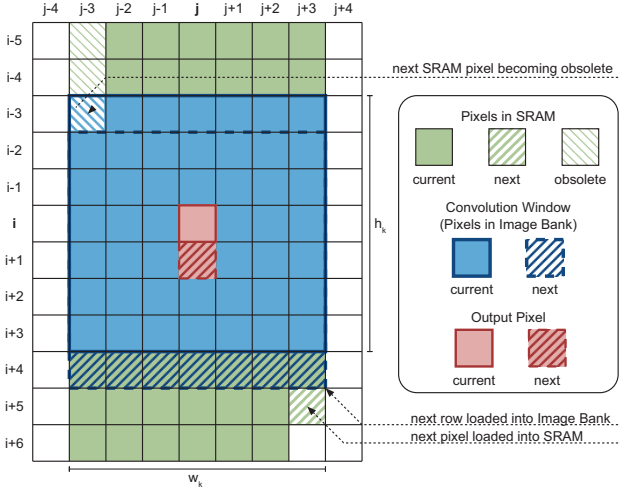}
    \caption[\optprint{Sliding window approach of the image memory.}]{\optprint{Sliding window approach of the image memory.}}
    \label{fig:memorg}
\end{figure}

\revAddC{We use the same sliding window approach developed in Cavigelli et al. \cite{Cavigelli2015}} and \figref{memorg} shows the implemented sliding window approach. To avoid shifting all images in the image memory to the left for the next column, the right most pixels are inserted at the position of the obsolete pixel, and the weights are shifted instead. To illustrate this, Equation \eqref{eq:conv22} shows the partial convolution for one pixel while the pixels are aligned to the actual column order and Equation \eqref{eq:conv32} shows it when the next column is processed and the weights need to be aligned. To indicate the partial sum, the Frobenius inner product formalism is used, where: $\langle\mathbf{A},\mathbf{B}\rangle_F=\sum_{i,j} a_{ij} b_{ij}$.

\begin{equation}\label{eq:conv22}
\tilde{o}(2,2)=\left \langle  
\begin{bmatrix}
    x_{11}       & x_{12} & x_{13} \\
    x_{21}       & x_{22} & x_{23}  \\
    x_{31}       & x_{32} & x_{33} 
\end{bmatrix},
\begin{bmatrix}
    w_{11}       & w_{12} & w_{13} \\
    w_{21}       & w_{22} & w_{23}  \\
    w_{31}       & w_{32} & w_{33} 
\end{bmatrix} \right \rangle_F
\end{equation}\vspace{0.2em}
\begin{equation}\label{eq:conv32}
\tilde{o}(3,2)=\left \langle  
\begin{bmatrix}
    x_{14}       & x_{12} & x_{13} \\
    x_{24}       & x_{22} & x_{23}  \\
    x_{34}       & x_{32} & x_{33} 
\end{bmatrix},
\begin{bmatrix}
    w_{13}       & w_{11} & w_{12} \\
    w_{23}       & w_{21} & w_{22}  \\
    w_{33}       & w_{31} & w_{32} 
\end{bmatrix} \right \rangle_F
\end{equation}

\revDel{As can be seen in Equation 3, when the pixels are shifted also the weights need to be shifted accordingly. Algebraically this weight shifting or permutation looks as follows and is implemented in hardware as a right circular shift registers.}\revAdd{Equation~\ref{eq:conv32} shows the operands as they are applied to the SoP units. The \nth{4} column which should be the most-right column is in the first column and also the other columns are shifted to the right, thus the weights also needs to be shifted to the right to obtain the correct result. The permutation in algebraic form is formulated in Equation \eqref{eq:conev32}:}
\begin{equation}\label{eq:conev32}
\tilde{o}(3,2)=\left \langle  
\begin{bmatrix}
    x_{14}       & x_{12} & x_{13} \\
    x_{24}       & x_{22} & x_{23}  \\
    x_{34}       & x_{32} & x_{33} 
\end{bmatrix}, 
\begin{bmatrix}
    w_{11}       & w_{12} & w_{13} \\
    w_{21}       & w_{22} & w_{23}  \\
    w_{31}       & w_{32} & w_{33} 
\end{bmatrix} \cdot P\right \rangle_F
\end{equation}

where $P=\begin{bmatrix}
    0       & 1 & 0 \\
    0       & 0 & 1  \\
    1       & 0 & 0
\end{bmatrix}$ is the permutation matrix

\subsection{BinaryConnect Approach}\label{sec:binaryconnect}

In this work we present a CNN accelerator based on BinaryConnect \cite{Courbariaux2015a}. With respect to an equivalent 12-bit version, the first major change in architecture are the weights which are reduced to a binary value $w_{k,n}\in\{-1,1\}$ and remapped by the following equation:
\begin{equation}
f \colon \{-1,1\} \to \{0,1\}, \, y \mapsto \begin{cases}
 0 & \; \text{if} \; z=-1 \\
 1 & \; \text{if} \; z=1 \\
\end{cases}
\end{equation}
The size of the filter bank decreases thus from $n_{ch}^2\cdot h_{k}^2 \cdot 12=37'632$ bit to $n_{ch}^2\cdot h_{k}^2\cdot 1=3'136$ bit in case of the 12-bit MAC architecture with $8\times 8$ channels and $7\times 7$ filters that we consider as baseline. The $12\times 12$-bit multipliers can be substituted by two's-complement operations and multiplexers, which reduce the "multiplier" and the adder tree size, as the products have a width of 12 bit instead of 24. The SoP is fed by a 12-bit and $7\times 7$ pixel sized image window and $7\times 7$ binary weights. Figure~\ref{fig:bararea} shows the impact on area while moving from 12-bit MACs to the binary connect architectures. Considering that with the 12-bit MAC implementation 40\% of the total total chip area is used for the filter bank and another 40\% are needed for the $12\times 12$-bit multipliers and the accumulating adder trees, this leads to a \revAddC{significant reduction} in  area cost and complexity. \revAdd{In fact the area of the conventional SoP unit could be reduced by \calc{1}{290.016/55.01666667}$\times$ and the filter bank by \calc{1}{309.168/20.73263889}$\times$ when moving from the Q2.9 to the binary version. The impact on the filter bank is straightforward as 12 times less bits need to be saved compared to the Q2.9, but also the SoP shrinks, as the 12$\times$12-bit multipliers are replaced with 2's complement \revAddC{operation units} and multiplexers and the adder tree needs to support a smaller dynamic range, thanks to the smaller products, since} the critical path is reduced as well. It is possible to reduce voltage while still keeping the same operating frequency and thus improving the energy efficiency even further.
\input{\plots{bar_area}}

\subsection{Latch-Based SCM}

An effective approach to optimize energy efficiency is to adapt the supply voltage of the architecture according to the performance requirements of the application. However, the potential of this approach is limited by the presence of SRAMs for implementation of image memory, which bounds the voltage scalability to \SI{.8}{\volt} (in 65nm CMOS technology). To overcome this limitation, we replace the SRAM-based image memory with a latch-based SCMs taking advantage of the area savings achieved through adoption of binary SoPs. 

\revAdd{Indeed, although SCMs are more expensive in terms of area (Figure~\ref{fig:bararea}), they are able to operate in the whole operating range of the technology (\SI{0.6}{\volt} - \SI{1.2}{\volt}) and they also feature significantly smaller read/write energy \cite{TemanSCM2016} at the same voltage. \revAdd{To reduce the area overhead of the SCMs and improve routability we propose a multi-banked implementation, where the image memory consists \revAddC{of a latch array organized in 6$\times$8 blocks of 128 rows of 12-bit values}, as described in Fig~\ref{fig:scmmemoryaccessscheme}. A pre-decoding logic, driven by the controller of the convolutional accelerator addresses the proper bank of the array every cycle, generating the local write and read enable signals, the related address fields, and propagating the input pixels to the banks and the current pixels to the SoP unit}. During a typical CNN execution, every cycle, 6 SCMs banks are read, and one is written, according to the image memory access pattern described in Fig~\ref{fig:memorg}.}

\revAdd{The SCMs are designed with a hierarchical clock gating and address/data silencing mechanisms as shown in Figure~\ref{fig:scmclockgates}, so that when a bank is not accessed the whole latch array consumes no dynamic power.} \revAddC{Every SCM block consists} of a 12\,bit $\times$ 128 rows array of latches, a data-in write path, and a read-out path. To meet the requirements of the application, the SCM banks are implemented with a two-ported, single-cycle latency architecture with input data and read address sampling. The write path includes data-in sampling registers, and a two-level clock gating scheme for minimizing the dynamic power of the clock path to the storage latches. The array write enable port drives the global clock gating cell, while the row clock gating cells are driven by the write address one-hot decoder. The readout path is implemented with a read address register with clock gating driven by a read enable signal, and a static multiplexer tree, which provides robust and low power operation, and enables dense and low congestion layout.

\revAdd{Thanks to this optimized architecture based on SCMs,} only up to 7 out of 48 banks of SCM banks consume dynamic power in every cycle, reducing power consumption of the memory by 3.25$\times$ at \SI{1.2}{\volt} with respect to a solution based on SRAMs \cite{Cavigelli2016}, while extending the functional range of the whole convolutional engine down to \SI{.6}{\volt} which is the voltage limit of the standard cells in UMC~65nm technology chosen for implementation \cite{MiaWallace}.

\begin{figure}
    \centering
    \includegraphics[width=.48\textwidth]{\figures{scm2}}
    \caption{Image memory architecture.}
    \label{fig:scmmemoryaccessscheme}
\end{figure}

\begin{figure}
    \centering
    \includegraphics[width=.48\textwidth]{\figures{scm1}}
    \caption{Block diagram of one SCM bank.}
    \label{fig:scmclockgates}
\end{figure}

\subsection{\revAddC{Considering I/O Power in Energy Efficiency}}
I/O power is a primary concern of convolutional accelerators, consuming even more than 30\% of the overall chip power \cite{Cavigelli2015a}. As we decrease the computational complexity by the binary approach, the I/O power gets even more critical. Fortunately, if the number of output channels is increased, more operations can be executed on the same data, which reduces the needed bandwidth and pad power consumption. The other advantage with having more SoP units on-chip is throughput which is formulated in \eqref{eq:throughput}:
\begin{equation} \label{eq:throughput}
\Theta = 2\cdot (n_{filt}^2 \cdot n_{ch})\cdot f
\end{equation}
With this in mind, we increased the number of input and output channels from $8\times 8$ to $16\times 16$ and $32\times 32$ which provides an ideal speed-up of throughput by $2\times$ and $4\times$, respectively.

\subsection{Support for Different Filter Sizes, Zero-Padding\revAddB{, Scaling and Biasing}}
Adapting filter size to the problem provides an effective way to improve the flexibility and energy efficiency of the accelerator when executing CNNs with different requirements. Although the simplest approach is to zero-pad the filters, this is not feasible in the presented binary connect architecture, as the value $0$ is mapped to $-1$. A more \revAdd{energy}-efficient approach tries to re-use parts of the architecture. We present an architecture where we re-use the binary multipliers for two $3\times 3$, two $5\times 5$ or one $7\times 7$ filters. In this work we limit the number of output channels per SoP unit to two as we are limited \revAdd{in output bandwidth.} \revDel{in the number of pins and can afford only 12 additional pins.} With respect to our baseline architecture, supporting only $7\times7$ filters, the number of binary operators and the weights per filter is increased from 49 to 50, such that two $5\times 5$ or one $7\times 7$ filter fits \revAdd{into} one \revAdd{SoP unit}. In case a filter size of $3\times 3$ or $5\times 5$ is used, the image from the image bank is mapped to the first 25 input image pixels, and the latter 25 and are finally accumulated in the adjusted adder tree, which is drawn in \figref{sop_addertree}. With this scheme, $n_{ch}\times 2n_{ch}$ channels for $3\times 3$ and $5\times 5$ filters can be calculated, which improves the maximum bandwidth and energy efficiency for these two cases. \revAdd{The unused 2's complement-and-multiplex operands \revAddC{(binary multipliers)} and the related part of the adder tree are silenced and clock-gated to reduce switching, therefore keeping the power dissipation as low as possible.}

\revAdd{To support also different kernel sizes, \revAddC{we provide the functionality to zero-pad the unused columns} from the image memory and the rows from the image bank \revAddC{instead of zeroing the weights which does not make sense with binary weights.}. This allows us to support \revAddC{kernels of size} $1\times 1$, $2\times 2$, $4\times 4$ and $6\times 6$ as well. The zero-padding is also used to add zeros to image borders: e.g. for a 7$\times$7 convolution the first 3 columns and first 3 rows of the \nth{4} column is preloaded. The 3 columns right to the initial pixel and the 3 rows on top of the pixel are zeroed the same way as described before and thus have not to be loaded onto the chip.}



\begin{figure}
    \centering
    \includegraphics[width=.40\textwidth]{\figures{sop_addertree_multi}}
    \caption{\revAddC{The adder tree in the SoP unit: Different colors are showing the data paths for $3\times 3$, $5\times 5$, $7\times 7$ kernels are indicated. The operands of the unused adders are silenced, but not indicated in the figure.}}
    \label{fig:sop_addertree}
    
\end{figure}

\revAddB{Finally, the system supports channel scaling and biasing which are common operations (e.g. in batch normalization layer) in neural networks which can be calculated efficiently.} As described in the previous section up to two output channels are calculated in parallel in every SoP unit, therefore the SoP saves also two scaling and two biasing values for these different output channels. As the feature maps are kept in maximum precision on-chip, the \revAddC{channel summers}' output Q7.9 fixed-point values, which are than multiplied with the Q2.9 formatted scaling factor and added to the Q2.9 bias and finally the Q10.18 output is resized with saturation and truncation to the initial Q2.9 format. With the interleaved data streaming, these operations are just needed once per cycle or twice when the number of output channels are doubled (e.g. $k=3\times 3$).

%% file: cnn_overview_detail.pdf_tex
\begingroup%
  \makeatletter%
  \providecommand\color[2][]{%
    \errmessage{(Inkscape) Color is used for the text in Inkscape, but the package 'color.sty' is not loaded}%
    \renewcommand\color[2][]{}%
  }%
  \providecommand\transparent[1]{%
    \errmessage{(Inkscape) Transparency is used (non-zero) for the text in Inkscape, but the package 'transparent.sty' is not loaded}%
    \renewcommand\transparent[1]{}%
  }%
  \providecommand\rotatebox[2]{#2}%
  \ifx\svgwidth\undefined%
    \setlength{\unitlength}{130.24416504bp}%
    \ifx\svgscale\undefined%
      \relax%
    \else%
      \setlength{\unitlength}{\unitlength * \real{\svgscale}}%
    \fi%
  \else%
    \setlength{\unitlength}{\svgwidth}%
  \fi%
  \global\let\svgwidth\undefined%
  \global\let\svgscale\undefined%
  \makeatother%
  \begin{picture}(1,0.71765254)%
    \put(0,0){\includegraphics[width=\unitlength]{\figures{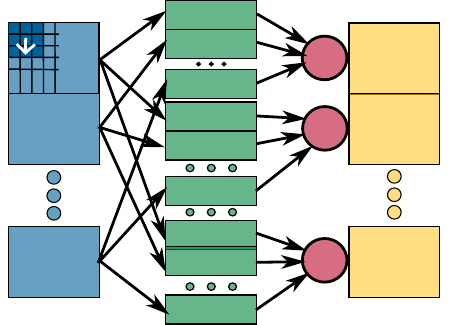}}}%
    \put(0.16774496,0.57535016){\color[rgb]{0,0,0}\makebox(0,0)[b]{\smash{$i_0$}}}%
    \put(0.11908973,0.41423485){\color[rgb]{0,0,0}\makebox(0,0)[b]{\smash{$i_1$}}}%
    \put(0.11908973,0.119754){\color[rgb]{0,0,0}\makebox(0,0)[b]{\smash{$i_{31}$}}}%
    \put(0.87196146,0.57024592){\color[rgb]{0,0,0}\makebox(0,0)[b]{\smash{$o_0$}}}%
    \put(0.87196146,0.41423485){\color[rgb]{0,0,0}\makebox(0,0)[b]{\smash{$o_1$}}}%
    \put(0.87196146,0.119754){\color[rgb]{0,0,0}\makebox(0,0)[b]{\smash{$o_{31}$}}}%
    \put(0.71771766,0.57471935){\color[rgb]{0,0,0}\makebox(0,0)[b]{\smash{$\Sigma$}}}%
    \put(0.4672271,0.67615903){\color[rgb]{0,0,0}\makebox(0,0)[b]{\smash{$i_0\ast w_{0,0}$}}}%
    \put(0.4672271,0.60626507){\color[rgb]{0,0,0}\makebox(0,0)[b]{\smash{$i_1\ast w_{0,1}$}}}%
    \put(0.4672271,0.52413806){\color[rgb]{0,0,0}\makebox(0,0)[b]{\smash{$i_{31}\ast w_{0,31}$}}}%
    \put(0.4672271,0.4459298){\color[rgb]{0,0,0}\makebox(0,0)[b]{\smash{$i_0\ast w_{1,0}$}}}%
    \put(0.4672271,0.38832404){\color[rgb]{0,0,0}\makebox(0,0)[b]{\smash{$i_1\ast w_{1,1}$}}}%
    \put(0.4672271,0.28162414){\color[rgb]{0,0,0}\makebox(0,0)[b]{\smash{$i_{31}\ast w_{1,31}$}}}%
    \put(0.4672271,0.18992883){\color[rgb]{0,0,0}\makebox(0,0)[b]{\smash{$i_0\ast w_{31,0}$}}}%
    \put(0.4672271,0.12003498){\color[rgb]{0,0,0}\makebox(0,0)[b]{\smash{$i_1\ast w_{31,1}$}}}%
    \put(0.4672271,0.02562321){\color[rgb]{0,0,0}\makebox(0,0)[b]{\smash{$i_{31}\ast w_{31,31}$}}}%
    \put(0.7184593,0.41896792){\color[rgb]{0,0,0}\makebox(0,0)[b]{\smash{$\Sigma$}}}%
    \put(0.7179305,0.12720826){\color[rgb]{0,0,0}\makebox(0,0)[b]{\smash{$\Sigma$}}}%
  \end{picture}%
\endgroup%

%% file: origami_top-v11.pdf_tex
\begingroup%
  \makeatletter%
  \providecommand\color[2][]{%
    \errmessage{(Inkscape) Color is used for the text in Inkscape, but the package 'color.sty' is not loaded}%
    \renewcommand\color[2][]{}%
  }%
  \providecommand\transparent[1]{%
    \errmessage{(Inkscape) Transparency is used (non-zero) for the text in Inkscape, but the package 'transparent.sty' is not loaded}%
    \renewcommand\transparent[1]{}%
  }%
  \providecommand\rotatebox[2]{#2}%
  \ifx\svgwidth\undefined%
    \setlength{\unitlength}{222.2697998bp}%
    \ifx\svgscale\undefined%
      \relax%
    \else%
      \setlength{\unitlength}{\unitlength * \real{\svgscale}}%
    \fi%
  \else%
    \setlength{\unitlength}{\svgwidth}%
  \fi%
  \global\let\svgwidth\undefined%
  \global\let\svgscale\undefined%
  \makeatother%
  \begin{picture}(1,0.43918936)%
    \put(0,0){\includegraphics[width=\unitlength]{\figures{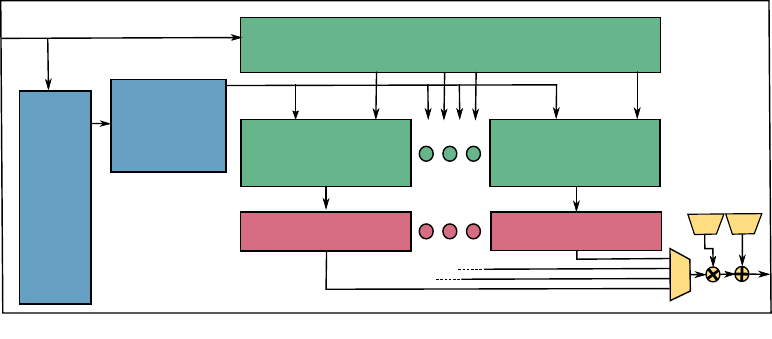}}}%
    \put(0.74411875,0.22851191){\color[rgb]{0,0,0}\makebox(0,0)[b]{\smash{SoP $31$ }}}%
    \put(0.06363565,0.1845556){\color[rgb]{0,0,0}\rotatebox{90}{\makebox(0,0)[b]{\smash{ImgMem}}}}%
    \put(0.41956663,0.22851191){\color[rgb]{0,0,0}\makebox(0,0)[b]{\smash{SoP $0$}}}%
    \put(0.59550374,0.37318573){\color[rgb]{0,0,0}\makebox(0,0)[b]{\smash{FilterBank $32\cdot 32\cdot 7\cdot 7\cdot 1$}}}%
    \put(0.21784967,0.28601918){\color[rgb]{0,0,0}\makebox(0,0)[b]{\smash{ImgBank}}}%
    \put(0.42254351,0.1270823){\color[rgb]{0,0,0}\makebox(0,0)[b]{\smash{ChSum $0$}}}%
    \put(0.7469513,0.12735454){\color[rgb]{0,0,0}\makebox(0,0)[b]{\smash{ChSum $31$}}}%
    \put(0.10020835,0.17726008){\color[rgb]{0,0,0}\rotatebox{89.75399007}{\makebox(0,0)[b]{\smash{$1024\cdot 7\cdot 12$}}}}%
    \put(0.21671942,0.24608655){\color[rgb]{0,0,0}\makebox(0,0)[b]{\smash{32$\cdot$$7^2$$\cdot$$12$}}}%
    \put(0.92657831,0.16496475){\color[rgb]{0,0,0}\rotatebox{90}{\makebox(0,0)[lb]{\smash{Scaling Factor}}}}%
    \put(0.97642403,0.16717843){\color[rgb]{0,0,0}\rotatebox{90}{\makebox(0,0)[lb]{\smash{Channel Bias}}}}%
  \end{picture}%
\endgroup%

%% file: timing2.pdf_tex
\begingroup%
  \makeatletter%
  \providecommand\color[2][]{%
    \errmessage{(Inkscape) Color is used for the text in Inkscape, but the package 'color.sty' is not loaded}%
    \renewcommand\color[2][]{}%
  }%
  \providecommand\transparent[1]{%
    \errmessage{(Inkscape) Transparency is used (non-zero) for the text in Inkscape, but the package 'transparent.sty' is not loaded}%
    \renewcommand\transparent[1]{}%
  }%
  \providecommand\rotatebox[2]{#2}%
  \ifx\svgwidth\undefined%
    \setlength{\unitlength}{406.77392578bp}%
    \ifx\svgscale\undefined%
      \relax%
    \else%
      \setlength{\unitlength}{\unitlength * \real{\svgscale}}%
    \fi%
  \else%
    \setlength{\unitlength}{\svgwidth}%
  \fi%
  \global\let\svgwidth\undefined%
  \global\let\svgscale\undefined%
  \makeatother%
  \begin{picture}(1,0.11796709)%
    \put(0,0){\includegraphics[width=\unitlength]{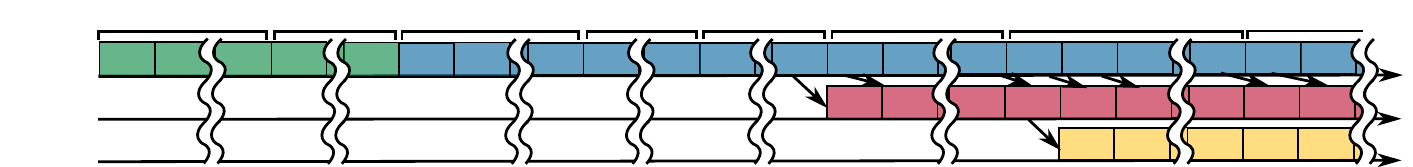}}%
    \put(0.06122838,0.06751463){\color[rgb]{0,0,0}\makebox(0,0)[rb]{\smash{in}}}%
    \put(0.06259115,0.03492029){\color[rgb]{0,0,0}\makebox(0,0)[rb]{\smash{SoP$k$}}}%
    \put(0.06227683,0.00741572){\color[rgb]{0,0,0}\makebox(0,0)[rb]{\smash{out}}}%
    \put(0.0904365,0.07094886){\color[rgb]{0,0,0}\makebox(0,0)[b]{\smash{$w_{0,0}$}}}%
    \put(0.1297351,0.07094886){\color[rgb]{0,0,0}\makebox(0,0)[b]{\smash{$w_{0,1}$}}}%
    \put(0.1727637,0.07094886){\color[rgb]{0,0,0}\makebox(0,0)[b]{\smash{$w_{0,31}$}}}%
    \put(0.21199153,0.07094886){\color[rgb]{0,0,0}\makebox(0,0)[b]{\smash{$w_{1,0}$}}}%
    \put(0.26321676,0.07069412){\color[rgb]{0,0,0}\makebox(0,0)[b]{\smash{$w_{31,31}$}}}%
    \put(0.30250878,0.07069412){\color[rgb]{0,0,0}\makebox(0,0)[b]{\smash{$i_0^{(0,0)}$}}}%
    \put(0.34163278,0.07069412){\color[rgb]{0,0,0}\makebox(0,0)[b]{\smash{$i_1^{(0,0)}$}}}%
    \put(0.39389862,0.07153816){\color[rgb]{0,0,0}\makebox(0,0)[b]{\smash{$i_{31}^{(0,0)}$}}}%
    \put(0.43315213,0.07153816){\color[rgb]{0,0,0}\makebox(0,0)[b]{\smash{$i_0^{(0,1)}$}}}%
    \put(0.47622624,0.07153816){\color[rgb]{0,0,0}\makebox(0,0)[b]{\smash{$i_{31}^{(0,31)}$}}}%
    \put(0.51545435,0.07153816){\color[rgb]{0,0,0}\makebox(0,0)[b]{\smash{$i_{31}^{(1,0)}$}}}%
    \put(0.56667887,0.07153816){\color[rgb]{0,0,0}\makebox(0,0)[b]{\smash{$i_0^{(6,6)}$}}}%
    \put(0.60536711,0.04078118){\color[rgb]{0,0,0}\makebox(0,0)[b]{\smash{$\tilde{o}_{k,0}^{(0,0)}$}}}%
    \put(0.60597091,0.07153816){\color[rgb]{0,0,0}\makebox(0,0)[b]{\smash{$i_1^{(6,6)}$}}}%
    \put(0.64449109,0.04078118){\color[rgb]{0,0,0}\makebox(0,0)[b]{\smash{$\tilde{o}_{k,1}^{(0,0)}$}}}%
    \put(0.64509491,0.07153816){\color[rgb]{0,0,0}\makebox(0,0)[b]{\smash{$i_2^{(6,6)}$}}}%
    \put(0.69337547,0.07201364){\color[rgb]{0,0,0}\makebox(0,0)[b]{\smash{$i_{31}^{(6,6)}$}}}%
    \put(0.73266749,0.07201364){\color[rgb]{0,0,0}\makebox(0,0)[b]{\smash{$i_0^{(6,7)}$}}}%
    \put(0.77179149,0.07201364){\color[rgb]{0,0,0}\makebox(0,0)[b]{\smash{$i_2^{(6,7)}$}}}%
    \put(0.73170461,0.04069298){\color[rgb]{0,0,0}\makebox(0,0)[b]{\smash{$\tilde{o}_{k,31}^{(0,0)}$}}}%
    \put(0.77082859,0.04069298){\color[rgb]{0,0,0}\makebox(0,0)[b]{\smash{$\tilde{o}_{k,0}^{(0,1)}$}}}%
    \put(0.69193699,0.04069298){\color[rgb]{0,0,0}\makebox(0,0)[b]{\smash{$\tilde{o}_{k,30}^{(0,0)}$}}}%
    \put(0.76956829,0.01140328){\color[rgb]{0,0,0}\makebox(0,0)[b]{\smash{$o_0^{(0,0)}$}}}%
    \put(0.80885173,0.01140325){\color[rgb]{0,0,0}\makebox(0,0)[b]{\smash{$o_1^{(0,0)}$}}}%
    \put(0.81107491,0.07208161){\color[rgb]{0,0,0}\makebox(0,0)[b]{\smash{$i_2^{(6,7)}$}}}%
    \put(0.81011205,0.04076096){\color[rgb]{0,0,0}\makebox(0,0)[b]{\smash{$\tilde{o}_{k,1}^{(0,1)}$}}}%
    \put(0.86070559,0.01140325){\color[rgb]{0,0,0}\makebox(0,0)[b]{\smash{$o_{31}^{(0,0)}$}}}%
    \put(0.86272477,0.07207374){\color[rgb]{0,0,0}\makebox(0,0)[b]{\smash{$i_0^{(6,8)}$}}}%
    \put(0.90184877,0.07207374){\color[rgb]{0,0,0}\makebox(0,0)[b]{\smash{$i_1^{(6,8)}$}}}%
    \put(0.86176192,0.04075309){\color[rgb]{0,0,0}\makebox(0,0)[b]{\smash{$\tilde{o}_{k,31}^{(0,1)}$}}}%
    \put(0.90088585,0.04075309){\color[rgb]{0,0,0}\makebox(0,0)[b]{\smash{$\tilde{o}_{k,0}^{(0,1)}$}}}%
    \put(0.94092182,0.07205604){\color[rgb]{0,0,0}\makebox(0,0)[b]{\smash{$i_2^{(6,8)}$}}}%
    \put(0.93995885,0.04073539){\color[rgb]{0,0,0}\makebox(0,0)[b]{\smash{$\tilde{o}_{k,0}^{(0,2)}$}}}%
    \put(0.89986804,0.01135208){\color[rgb]{0,0,0}\makebox(0,0)[b]{\smash{$o_0^{(0,1)}$}}}%
    \put(0.93915147,0.01135205){\color[rgb]{0,0,0}\makebox(0,0)[b]{\smash{$o_1^{(0,1)}$}}}%
    \put(0.99366445,0){\color[rgb]{0,0,0}\makebox(0,0)[lb]{\smash{t}}}%
    \put(0.99366445,0.02945657){\color[rgb]{0,0,0}\makebox(0,0)[lb]{\smash{t}}}%
    \put(0.99366445,0.06037574){\color[rgb]{0,0,0}\makebox(0,0)[lb]{\smash{t}}}%
    \put(0.12762441,0.11138052){\color[rgb]{0,0,0}\makebox(0,0)[b]{\smash{load weights}}}%
    \put(0.12765844,0.10019002){\color[rgb]{0,0,0}\makebox(0,0)[b]{\smash{$n_{in}=0$}}}%
    \put(0.23905765,0.11138051){\color[rgb]{0,0,0}\makebox(0,0)[b]{\smash{load weights}}}%
    \put(0.23909168,0.10019002){\color[rgb]{0,0,0}\makebox(0,0)[b]{\smash{$n_{in}=1,...,31$}}}%
    \put(0.3468085,0.11138051){\color[rgb]{0,0,0}\makebox(0,0)[b]{\smash{preload $\forall n_{in}$ }}}%
    \put(0.34684253,0.10019002){\color[rgb]{0,0,0}\makebox(0,0)[b]{\smash{pixel $(0,0)$}}}%
    \put(0.45289557,0.11116156){\color[rgb]{0,0,0}\makebox(0,0)[b]{\smash{preload $\forall n_{in}$ }}}%
    \put(0.45292961,0.09997107){\color[rgb]{0,0,0}\makebox(0,0)[b]{\smash{column $0$}}}%
    \put(0.53837656,0.11076402){\color[rgb]{0,0,0}\makebox(0,0)[b]{\smash{load col 1-6 \& }}}%
    \put(0.53841059,0.09957352){\color[rgb]{0,0,0}\makebox(0,0)[b]{\smash{pixels (0-6,6)}}}%
    \put(0.65151826,0.11116157){\color[rgb]{0,0,0}\makebox(0,0)[b]{\smash{load pixel (6,6) }}}%
    \put(0.65155229,0.09997107){\color[rgb]{0,0,0}\makebox(0,0)[b]{\smash{calculate part. sums}}}%
    \put(0.80140849,0.11116157){\color[rgb]{0,0,0}\makebox(0,0)[b]{\smash{load p. (6,7), calc. part. sums }}}%
    \put(0.80144252,0.09997107){\color[rgb]{0,0,0}\makebox(0,0)[b]{\smash{stream ouput feat. maps}}}%
    \put(0.93138485,0.09997107){\color[rgb]{0,0,0}\makebox(0,0)[b]{\smash{...}}}%
  \end{picture}%
\endgroup%

%% file: bar_area.tex
\begin{figure}[ht]
\centering
\begin{tikzpicture}[scale=0.75]

\pgfplotsset{
    calculate offset/.code={
        \pgfkeys{/pgf/fpu=true,/pgf/fpu/output format=fixed}
        \pgfmathsetmacro\testmacro{(\pgfplotspointmeta *10^\pgfplots@data@scale@trafo@EXPONENT@y)*\pgfplots@y@veclength)}
        \pgfkeys{/pgf/fpu=false}
    },
    every node near coord/.style={
        /pgfplots/calculate offset,
        yshift=-\testmacro
    },
}
\begin{axis}[ legend style={draw=none},x tick label style={ /pgf/number format/1000 sep=}, ylabel={Area {[kGE]}}, enlargelimits=0.15,  ybar, ytick={0,100,200,300,400,500}, 
xtick=data, legend style={at={(0.5,-0.2), draw=none}, anchor=north,legend columns=-1},
        xticklabels = {
            SoP,
            ImgBank,
            FilterBank,
            ImgMem,
            Others
        },
    bar width=0.15, nodes near coords={}, width=0.65\textwidth, ymin=62.5, height=150,
        ] 

\addplot[cieee3!20!black,fill=cieee3!60!white] table[x index=0,y index=3,restrict expr to domain={\thisrow{show}}{1:1}] {\tables{area_breakdown.csv}};

\addplot[cieee0!20!black,fill=cieee0!60!white] table[x index=0,y index=4,restrict expr to domain={\thisrow{show}}{1:1}] {\tables{area_breakdown.csv}};

\addplot[cieee1!20!black,fill=cieee1!60!white] table[x index=0,y index=5,restrict expr to domain={\thisrow{show}}{1:1}] {\tables{area_breakdown.csv}};

\addplot[cieee6!20!black,fill=cieee6!60!white] table[x index=0,y index=6,restrict expr to domain={\thisrow{show}}{1:1}] {\tables{area_breakdown.csv}};
  





\legend{{Q2.9-8x8}, Binary-8x8, Binary-16x16, Binary-32x32} 

\end{axis} \end{tikzpicture}
\caption{Area Breakdown for fixed-point and several binary architectures.}
\label{fig:bararea}
\end{figure}

%% file: 4_results.tex
\section{Results}
\subsection{Computational Complexity and Energy Efficiency Measure}
Research in the field of deep learning is done on a large variety of systems, such that platform-independent performance metrics are needed. For computational complexity analysis the total number of multiplications and additions has been used in other publications~\cite{Chen:2014:DSH:2654822.2541967,Farabet:2011_NeuFlow,Pham:2012_Neuflow,Cavigelli2015}. For a CNN layer with $n_{in}$ input channels and $n_{out}$ output channels, a filter kernel size of $h_k\times w_k$, and an input size of $h_{im}\times w_{im}$, the computational complexity to process one frame can be calculated as follows:
\begin{equation}\label{eq:complex}
    \#Op = 2n_{out}n_{in}h_kw_k(h_{in}-h_k+1)(w_{in}-h_k+1)
\end{equation}
The factor of 2 considers additions and multiplications as separate arithmetic operations (Op), while the rest of the equation calculates the number of \revAdd{multiply-accumulate operations} MACs. The two latter factors \revAddC{$(h_{in}-h_k+1)$ and $(w_{in}-h_k+1)$} \revAdd{are the height and width of the output channels including the reduction at the border in case no zero-padding was applied.}\revDel{account for the fact that the output image is reduced at the border by $(h_k-1)$ and $(w_k-1)$, respectively.} Memory \revAdd{accesses} are not counted as additional operations. The formula does not take into account the amount of operations executed when applying zero-padding. In the following evaluation, we will use the following metrics:
\begin{itemize}
\item Throughput $\Theta = (\text{\#Op based on \eqref{eq:complex}})/t\ {[\SI{}{\giga Op/\second}]}$
\item Peak Throughput: Theoretically reachable throughput. This does not take into account idling, cache misses, etc.
\item Energy Efficiency $H_E=\Theta/P\ {[\SI{}{\tera Op/\second/\watt}]}$
\item Area Efficiency $H_A=\Theta/A\ {[\SI{}{\giga Op/\second/\mega GE}]}$
\end{itemize}

\label{sec:realeff} 
\revAddB{Furthermore, we will introduce} some efficiency metrics to allow for realistic performance estimates, as CNN layers have varying numbers of input and output channels and image sizes vary from layer to layer.

\begin{equation}
    \Theta_{real} = \Theta_{peak}\cdot \prod_{i} \eta_{i} 
\end{equation}

\textbf{Tiling}: \new{The number of rows are limited by the image window memory, which accommodates  $h_{max}\cdot n_{ch, in}$ words of $w_k\cdot12$ bit, storing a maximum of $h_{max}$ rows per input channel. In case the full image height does not fit into the memory, it can be  split into several image tiles which are then processed consecutively. The penalty are the $(h_k-1)$ rows by which the tiles need to vertically overlap and thus are loaded twice. The impact on throughput can be determined by the tiling efficiency 
}
\begin{equation}
    \eta_{tile}=\frac{h_{im}}{h_{im}+\big(\ceil*{\frac{h_{im}}{h_{max}}}-1\big)(h_k-1)}. 
\end{equation}

    \textbf{(Input) Channel Idling:} \new{The number of output and input channels usually does not correspond to the number of output and input channels processed in parallel by this core. The output and input channels are partitioned into blocks of $n_{ch} \times n_{ch}$. Then the outputs of these blocks have to be summed up pixel-wise outside the accelerator.} 

\new{In the first few layers, the number of input channels $n_{in}$ can be smaller than the number of output channels $n_{out}$. In this case, the output bandwidth is limiting the input bandwidth by a factor of $\eta_{chIdle}$,}
\begin{equation}
\eta_{chIdle}=\frac{n_{in}}{n_{out}}.
\end{equation}
\new{Note that this factor only impacts throughput, not energy efficiency. Using less than the maximum available number of input channels only results in more cycles being spent idling, during which only a negligible amount of energy (mainly leakage) is dissipated.}

\new{\textbf{Border Considerations:} To calculate one pixel of an output channel, at least $h_k^2$ pixels of each input channel are needed. This leads to a reduction of $\frac{1}{2}(h_k-1)$ pixels on each side. While in some cases this is acceptable, many and particularly deep CNNs perform zero-padding to keep a constant image size, adding an all-zero halo around the image. \revAdd{In case of zero-padding, $\frac{h_k-1}{2}$ columns need to be pre-loaded, this introduces latency, but does not increase idleness as the same number of columns need to be processed after the last column where in the meantime the first columns of the next image can be preloaded to the image and therefore $\eta_{border}=1$.}\revDel{YodaNN supports on-the-fly zero-padding, such that every pixel that is streamed into the device is immediately processed resulting in no idle time and $\eta_{border}=1$.} For non-zero padded \revAdd{layers}, the efficiency is reduced by the factor}
\begin{equation}
\eta_{border,non\text{-}zero\text{-}padded}=\frac{h_k-1}{w_{im}}\cdot \frac{h_k-1}{h_{im}}.
\end{equation}

\subsection{\revAdd{Experimental Setup}}
\revAdd{To evaluate the performance and energy metrics of the proposed architecture and to verify the correctness of the generated results, we developed a testbench, which generates the control signals of the chip, reads the filters and the input images from a raw file, and streams the data to the chip. The output is monitored and compared to the expected output feature maps which are read from a file, too. To calculate the expected responses we have implemented a bit-true quantized spatial convolution layer in Torch which acts as a golden model. 
The power results are based on post place \& route results of the design. The design was synthesized with Synopsys Design Compiler J-2014.09-SP4, while place and route was performed with Cadence Innovus 15.2. The UMC 65nm standard cell libraries used for implementation were characterized using Cadence Liberate 12.14 in the voltage range \SI{.6}{\volt} - \SI{1.2}{\volt}, and in the typical process corner at the temperature of \SI{25}{\celsius}. The power simulations were performed with Synopsys PrimePower 2012.12, based on Value Change Dump (VCD) files extracted from simulations of real-life workloads running on the post place and route netlist of the design. These simulations were done with the neural network presented in \cite{Cavigelli2015a} on the Stanford backgrounds data set \cite{Gould2009} (715 images, $320\times 240$ RGB, scene-labeling for various outdoor scenes), where every pixel is assigned with one of 8 classes: sky, tree, road, grass, water, building, mountain and foreground object. The I/O power was approximated by power measurements on chips of the same technology \cite{Cavigelli2016} and scaled to the actual operating frequency of YodaNN.}

\begin{figure}
    \centering
    \includegraphics[width=0.45\textwidth]{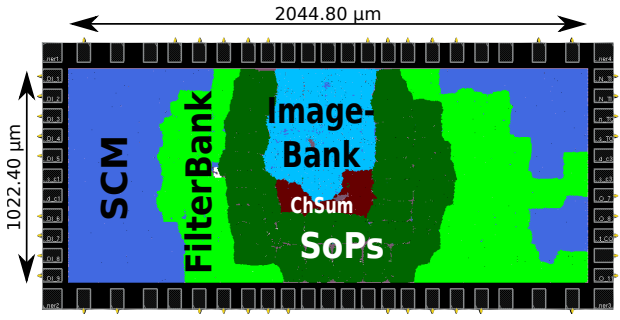}
    \caption{Floorplan of YodaNN with a \SI{9.2}{\kibi\byte} SCM memory computing 32 output channels in parallel.}
   \label{fig:chip_layout}
\end{figure}

\revAddB{The final floorplan of YodaNN is shown in \figref{chip_layout}. The area is split mainly among the SCM memory with 480\,kGE, the binary weights filter bank with 333\,kGE, the SoP units with 215\,kGE, the image bank with 123\,kGE and the area distribution is drawn in \figref{bararea}. The core area is 1.3\,MGE (\SI{1.9}{\milli\meter^2}). The chip runs at a maximum frequency of 480\,MHz\,@\,1.2\,V and 27.5\,MHz@0.6\,V.}

%
\subsection{Fixed-Point vs. YodaNN}
In this section, we compare a fixed-point baseline implementation with a binary version with fixed filter kernel size of $7\times 7$ and $8\times 8$ channels including an SRAM for input image storage. The results are summarized in Table~\ref{tbl:origvsbin}. The reduced arithmetic complexity and the replacement of the SRAM by a latch-based memory shortened the critical path delay. Three pipeline stages between the memory and the channel summers were used in the fixed-point baseline version \revAdd{could be reduced to one pipeline stage}. The peak throughput could still be increased from \SI{348}{\giga Op/\second} to \SI{377}{\giga Op/\second} at a core voltage of \SI{1.2}{\volt} and the core power was reduced by \calcsip{0}{(1-39.3/185)*100} to \SI{39}{\milli\watt}, which leads to a $\calc{1}{9.61/1.88}\times$ better core energy efficiency and $\calc{1}{631/487}\times$ better core area efficiency. As UMC 65nm technology SRAMs fail below \SI{0.8}{\volt}, we can get even better results by reducing the supply voltage to \SI{0.6}{\volt} thanks to our SCM implementation. Although the peak throughput drops to \SI{15}{\giga Op/\second}, the core power consumption is reduced to \SI{260}{\micro\watt}, and core energy efficiency rises to \SI{59}{\tera Op/\second/\watt}, which is an improvement of \calc{1}{58.56/5.04}$\times$ compared to the fixed-point architecture at \SI{0.8}{\volt}.

\begin{table}[ht] 
	\caption{Fixed-Point Q2.9 vs. Binary Architecture 8$\times$8}
	\label{tbl:origvsbin}
	\centering
	\footnotesize
	\setlength\tabcolsep{5.5pt}
	\begin{threeparttable}
	\begin{tabularx}{1.0\columnwidth}{l|rr|rr|r}
\toprule
Architecture                 & Q2.9\tnote{a}     & Bin.                     & Q2.9\tnote{a}   & Bin.   & Bin.  \\
Supply (V)                              & 1.2    & 1.2                      & 0.8   & 0.8    & 0.6   \\ \midrule
Peak Throughput (GOp/s)                 & 348    & 377                      & 131    & 149    & 15    \\
Avg. Power Core (mW)                    & 185  & 39                         & 31  & 5.1    & 0.26  \\
Avg. Power Device (mW)                  & 580   & 434                       & 143   & 162    & 15.54    \\
Core Area (MGE)                         & 0.72   & 0.60                     & 0.72  & 0.60   & 0.60  \\ \midrule
\multicolumn{6}{c}{Efficiency metrics}           \\
Energy Core (TOp/s/W)                   & 1.88  & 9.61                      & 4.26  & 29.05  & 58.56 \\
Energy Device (TOp/s/W)                  & 0.60      & 0.87                 & 0.89   & 0.92   & 0.98  \\
Area Core (GOp/s/MGE)                   & 487 & 631                         & 183 & 247 & 25 \\
Area Dev. (GOp/s/MGE)                   & 161  & 175                        & 61    & 69  & 7.0  \\ \bottomrule
\end{tabularx}
 \begin{tablenotes} 
    	\item[a]{A fixed-point version with SRAM is used as baseline comparison and 8$\times$8 channels and 7$\times$7 filters.}
    \end{tablenotes} 
\end{threeparttable}
\end{table}

\begin{figure}[ht]
\centering
\begin{tikzpicture}[scale=0.75]
\tikzset{mark options={solid, mark size=3, line width=1pt}} %
	\pgfplotsset{
 	 xmin=0.55, xmax=1.249
	}
	\begin{axis}[width=1.33\columnwidth-1.5cm, height=7cm, 
		xlabel={Supply Voltage $V_\text{core}$ {[V]}},ylabel={Core Energy Efficiency {[TOp/s/W]}}, grid=major, ymin=0, ymax=80, ytick={25,50,75,100}]

\addplot[color=blue,mark=*] table [x=supply, y=eneffc, col sep=space, restrict expr to domain={\thisrow{channels}}{8:8}, restrict expr to domain={\thisrow{type}}{12:12}]{\tables{results_mf.csv}}; \label{plotEffSCM:OrigEnEff}
    \addplot[color=red,mark=*] table [x=supply, y=eneffc, col sep=space, restrict expr to domain={\thisrow{channels}}{32:32}, restrict expr to domain={\thisrow{type}}{3:3}, restrict expr to domain={\thisrow{kernel}}{7:7}]{\tables{results_mf.csv}}; \label{plotEffSCM:SCMEnEff}
    
    	\end{axis}


		\begin{axis}[width=1.33\columnwidth-1.5cm, height=7cm,  xlabel={Voltage {[V]}},ylabel={Throughput {[GOp/s]}}, axis y line*=right, axis x line=none, grid=none, ymin=0, ymax=1600, ytick={500,1000,1500},
		]

  \addplot[color=blue,densely dashed, thick, mark=diamond*,unbounded coords=discard] table [x=supply, y=throughput, col sep=space, restrict expr to domain={\thisrow{channels}}{8:8}, restrict expr to domain={\thisrow{type}}{12:12}]{\tables{results_mf.csv}}; \label{plotEffSCM:OrigThrough}
  
\addplot[color=red,densely dashed, thick, mark=diamond*,] table [x=supply, y=throughput, col sep=space, restrict expr to domain={\thisrow{channels}}{32:32}, restrict expr to domain={\thisrow{type}}{3:3}, restrict expr to domain={\thisrow{kernel}}{7:7}]{\tables{results_mf.csv}};\label{plotEffSCM:SCMThrough}

	\end{axis}%

\end{tikzpicture}\\
\begin{scriptsize}	
		Core Energy Eff. Q2.9/8$\times$8/SRAM (\ref{plotEffSCM:OrigEnEff}), Bin./32$\times$32/SCM (\ref{plotEffSCM:SCMEnEff}), \\Throughput Q2.9/8$\times$8/SRAM (\ref{plotEffSCM:OrigThrough}), Bin./32$\times$32/SCM (\ref{plotEffSCM:SCMThrough})
	\end{scriptsize}

\caption{Comparison of core energy efficiency and throughput for the baseline architecture (fixed-point Q2.9, SRAM, 8$\times$8 channels, fixed 7$\times$7 filters) with final YodaNN (binary, SCM, 32$\times$32 channels, supporting several filters).}
\label{fig:plotEffSCM}
\end{figure}
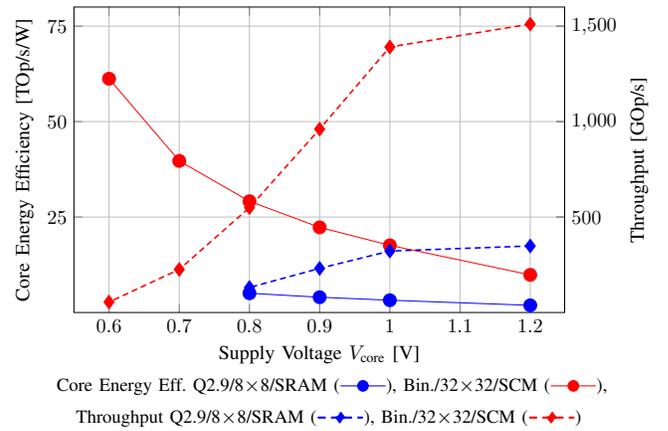

 \figref{plotEffSCM} shows the throughput and energy efficiency of YodaNN with respect to the baseline architecture for different voltage supplies, while \figref{barpower} shows the breakdown of the core power an the operating frequency of 400\,MHz. \revAdd{Comparing the two 8$\times$8 channels variants (fixed-point and binary weights), the power consumption was reduced from \SI{185}{\milli\watt} to \SI{39}{\milli\watt}, where the power could be reduced by $\calc{1}{13.02/3.7076}\times$ in the SCM, $\calc{1}{109.10/22.6412}\times$ in the SoP units and $\calc{0}{15.80/0.506}\times$ in the filter bank.} Although the power consumption of the core increases by $3.32\times$ when moving from $8\times 8$ to $32\times 32$ channels, the throughput increases by $4\times$, improving energy efficiency by 20\%. Moreover, taking advantage of more parallelism, voltage and frequency scaling can be exploited to improve energy efficiency for a target throughput. The support for different kernel sizes significantly improves the flexibility of the YodaNN architecture, but increases the core area by 11.2\%, and the core power by 38\% with respect to a binary design supporting 7$\times$7 kernels only. \revAdd{The Scale-Bias unit occupies another \SI{2.5}{\kilo GE} area and consumes \SI{0.4}{\milli\watt} at a supply voltage of \SI{1.2}{\volt} and a operating frequency of \SI{480}{\mega\hertz}.} \revAddC{When I/O power is considered, increasing the number of channels is more beneficial}, since we can increase the throughput while the total device power does not increase at the same rate. We estimate a fixed contribution of \SI{328}{\milli\watt} for the the I/O power at \SI{400}{MHz}. Table~\ref{tab:devenfilt} provides an overview of the device energy efficiency for different filter kernel sizes at \SI{1.2}{\volt} core and \SI{1.8}{\volt} pad supply. The device energy efficiency raises from \SI{856}{GOps/s/W} in the 8$\times$8 architecture to \SI{1611}{} in the 16$\times$16 and to 2756 in the 32$\times$32.

\input{\plots{power_breakdownAll.tex}}

\begin{table}[ht] %
\caption{Device Energy Efficiency for Different Filters and Architectures}
\label{tab:devenfilt}
	\centering
	\footnotesize
	\setlength\tabcolsep{5.5pt}
	\begin{tabularx}{1.0\columnwidth}{r|rrrrr|l}
\toprule
Archit. & Q2.9                  & 8$\times$8  & 16$\times$16 & 32$\times$32  & $32^2$ (fixed)     &        \\ \midrule
7$\times$7          &    600                                       & 856  &  1611 & 2756   & 3001 & [GOp/s/W] \\ 
5$\times$5          &                                              & 611  &  1170 & 2107   &      & [GOp/s/W] \\ 
3$\times$3          &                                              & 230  &  452  &  859   &      & [GOp/s/W] \\ 
\bottomrule
\end{tabularx}

\end{table}

\subsection{Real Applications}\label{sec:real}
For a comparison based on real-life CNNs, we have selected several state-of-the-art networks which exploit binary weights. This includes the CNNs from the BinaryConnect paper for Cifar-10 and SVHN \cite{Courbariaux2015a}, and the well-known networks VGG-13, VGG-19 \cite{Simonyan2014}, ResNet-18, ResNet-34 \cite{He2015}, and AlexNet~\cite{Krizhevsky2012}, which were successfully implemented with binary weights by Rastegari et al.~\cite{Rastegari2016} (not XNOR-net). The layer configurations and the related metrics are summarized in Table~\ref{tbl:realapp}. As described in Section~\ref{sec:dataflow}, the layers are split into blocks of $n_{in}\times n_{out} =32\times 32$ channels in case of a kernel size of $h_k^2=7^2$ and $n_{in}\times n_{out} =32\times 64$ elsewhere. \revAdd{The first layers have a high idle rate, but the silenced SoP units consume roughly no power. To account for this we introduce $\tilde{P}_{real}=P_{eff}/P_{max}$ which is calculated.} The first layer of AlexNet uses $11\times 11$ filters and needs to be split into \revAdd{smaller kernels. We split them into 2 filters of $6\times 6$ (top-left, bottom-right) and 2 filters of $5\times 5$ (bottom-left, top-right), where the center pixel is overlapped by both $6\times 6$ kernels. By choosing the value for the overlapping weight appropriately, it is possible to prevent the need of additional $1\times 1$ convolutions: if the original weight is 1, the overlapping weight of both $6\times 6$ kernels are chosen to be $1$, otherwise $-1$ is assigned to one of them and $1$ to the other. Instead of $1\times 1$ convolutions, just the sum of the identities of all input channels needs to be subtracted. The summing of the contributions and subtracting of the identities is done off-chip.}
\input{\tables{realapplication.tex}}

Table~\ref{tab:realnetworksenergy} gives an overview of the energy efficiency, throughput, actual frame rate and total energy consumption for calculating the convolutions, including channel biasing and scaling in the energy-optimal configuration (at \SI{0.6}{\volt}). Table~\ref{tab:realnetworkthroughput} shows the same metrics and CNNs for the high-throughput setting at \SI{1.2}{\volt}. 
It can be noticed that in the energy-optimal operating point, the achieved throughput is about half of the maximum possible throughput of \SI{55}{GOp/s} for most of the listed CNNs. This can be attributed to the smaller-than-optimal filter size of $3\times 3$, which is frequently used and limits the throughput to about \SI{20}{GOp/s}. However, note that the impact on peak energy-efficiency is only minimal with \SI{59.20}{} instead of \SI{61.23}{GOp/s/W}. 

The average energy efficiency of the different networks is within the range from \SI{48.1}{} to \SI{56.7}{TOp/s/W}, except for AlexNet which reaches \revAdd{\SI{14.1}{TOp/s/W}} due to the dominant first layer which requires a high computational effort while leaving the accelerator idling for a large share of the cycles because of the small number of input channels. The fourth column in tables~\ref{tab:realnetworksenergy} and \ref{tab:realnetworkthroughput} shows the frame rate which can be processed by YodaNN excluding the fully connected layers and the chip configuration. In the throughput optimal case, the achieved frame rate is between \SI{13.3}{} (for VGG-19) and \SI{1428}{FPS} (for the BinaryConnect-SVHN network)\revDel{ is achieved} with a chip power of just \SI{153}{\milli\watt}. In the maximum energy efficiency \revAdd{corner} YodaNN achieves a frame rate \revAdd{between \SI{0.5}{} and \SI{53.2}{FPS}} at a power of \SI{895}{\micro\watt}. 

\begin{table}[ht] %
\caption{Overview of Several Networks in an Energy Optimal Use Case ($V_\text{core}=\SI{0.6}{\volt}$) on a YodaNN Accelerator}
\label{tab:realnetworksenergy}
	\centering
	\footnotesize
	\setlength\tabcolsep{5.5pt}
	\begin{tabularx}{0.95\columnwidth}{l|r|rrrrr}
\toprule 
\multirow{2}{*}{Network}     & img size&Avg. EnEff & $\bar{\Theta}$ & $\Theta$   & Energy \\
            & $h_{in}$$\times$$w_{in}$ & TOp/s/W & GOp/s & FPS & \textmu J \\ \midrule
BC-Cifar-10 & 32$\times$32 & 56.7 & 19.1 & \calc{1}{1000/63.2}   & 21  \\
BC-SVHN     & 32$\times$32 & 50.6 & 16.5 & \calc{1}{1000/18.8}   & 6   \\
AlexNet     & 224$\times$224 & 14.1 & 3.3  & \calc{1}{1000/1963.7} & 352 \\
ResNet-18   & 224$\times$224 & 48.1 & 16.2 & \calc{1}{1000/221.5}  &  73 \\
ResNet-34   & 224$\times$224 & 52.5 & 17.8 & \calc{1}{1000/406.1} & 136 \\
VGG-13      & 224$\times$224 & 54.3 & 18.2 & \calc{1}{1000/1239.0} & 398 \\
VGG-19      & 224$\times$224 & 55.9 & 18.9 & \calc{1}{1000/2069.2} & 684  \\
\bottomrule
\end{tabularx}

\end{table}
\begin{table}[ht] %
\caption{Overview of Several Networks in a Throughput Optimal Use Case ($V_\text{core}=\SI{1.2}{\volt}$) on a YodaNN Accelerator}
\label{tab:realnetworkthroughput}
	\centering
	\footnotesize
	\setlength\tabcolsep{5.5pt}
	\begin{tabularx}{1\columnwidth}{l|r|rrrrr}
\toprule
\multirow{2}{*}{Network} & img size     & Avg. EnEff & $\bar{\Theta}$ & $\Theta$   & Energy \\
            & $h_{in}$$\times$$w_{in}$ & TOp/s/W & GOp/s & FPS & \textmu J \\ \midrule
BC-Cifar-10 & 32$\times$32 & 8.6 & 525.4 & \calc{1}{1000/2.3} & 137  \\
BC-SVHN     & 32$\times$32 & 7.7 & 454.4 & \calc{1}{1000/0.7} & 36   \\
AlexNet     & 224$\times$224 & 2.2 & 89.9 & \calc{1}{1000/71.2} & 2244 \\
ResNet-18   & 224$\times$224 & 7.3 & 446.4 & \calc{1}{1000/8} & 478 \\
ResNet-34   & 224$\times$224 & 8.0 & 489.5 & \calc{1}{1000/14.7} & 889 \\
VGG-13      & 224$\times$224 & 8.3 & 501.8 & \calc{1}{1000/44.6} & 2609 \\
VGG-19      & 224$\times$224 & 8.5 & 519.8 & \calc{1}{1000/75.1} & 4482  \\
\bottomrule
\end{tabularx}
\end{table}







\input{\plots{soa.tex}}

\footnotetext{\revAdd{The $11\times 11$ kernels are split into 2 $6\times 6$ and 2 $5\times 5$ kernels as described in Section \ref{sec:real}.} \revDel{2 $5\times 5$ and 4 $3\times 3$ filters} \label{footnotehk11}}

\subsection{Comparison with State-of-the-Art}
In Section~\ref{sec:relatedwork}, \revAddC{the literature from several software and architectural approaches have been described}. The $32\times 32$ channel YodaNN is able to reach a peak throughput of \SI{1.5}{\tera Op/\second} which outperforms NINEX \cite{park2016} by a factor of \calc{1}{1510/569}. In core energy efficiency the design outperforms k-Brain, NINEX by \calc{0}{9862/1930}$\times$ and more. If the supply voltage is reduced to \SI{0.6}{\volt}, the throughput decreases to \SI{55}{\giga Op/\second} but the energy efficiency rises to \SI{61.2}{\tera Op/\second}, which is more than an order-of-magnitude improvement over the previously reported results \cite{kbrain2015, park2016, Jaehyeong2016}. 
The presented architecture also outperforms the compressed neural network accelerator EIE in terms of energy efficiency by \calc{0}{61234/5000}$\times$ and in terms of area efficiency by \calc{0}{1135.3/41.2}$\times$, even though they assume a very high degree of sparsity with 97\% zeros \cite{Han2016b}. \revAdd{\figref{soaAreaEnergy} gives a quantitative comparison of the state-of-the-art in energy efficiency and area efficiency. For the sweep of voltages between \SI{1.2}{\volt} and \SI{0.6}{\volt}, YodaNN builds a clear pareto front over the state of the art.}


%% file: power_breakdownAll.tex
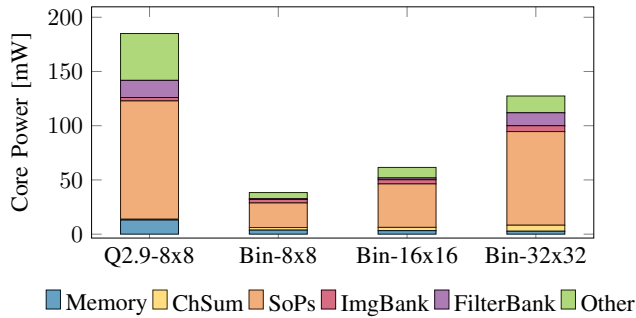
\begin{figure}[ht]
\centering
\pgfplotsset{width=9cm, height=5cm}
\begin{tikzpicture}[scale=0.9]

\begin{axis}[legend style={draw=none}, ybar stacked, xtick=data,
        xticklabels = {
            Q2.9-8x8,
            Bin-8x8,
            Bin-16x16,
            Bin-32x32,
            32x32N
        },
    bar width=0.45, nodes near coords={}, enlargelimits=0.15, ymin= 21,   ylabel={Core Power {[mW]}}, 
    legend style={at={(0.5,-0.2)}, anchor=north,legend columns=-1},
        ] 
\addplot[cieee0!20!black,fill=cieee0!60!white] table[x index=0,y index=4,restrict expr to domain={\thisrow{show}}{1:1}] {\tables{power_breakdownAll.csv}};
\addplot[cieee1!20!black,fill=cieee1!60!white] table[x index=0,y index=5,restrict expr to domain={\thisrow{show}}{1:1}] {\tables{power_breakdownAll.csv}};
\addplot[cieee2!20!black,fill=cieee2!60!white] table[x index=0,y index=6,restrict expr to domain={\thisrow{show}}{1:1}] {\tables{power_breakdownAll.csv}};
\addplot[cieee3!20!black,fill=cieee3!60!white] table[x index=0,y index=7,restrict expr to domain={\thisrow{show}}{1:1}] {\tables{power_breakdownAll.csv}};
\addplot[cieee4!20!black,fill=cieee4!60!white] table[x index=0,y index=8,restrict expr to domain={\thisrow{show}}{1:1}] {\tables{power_breakdownAll.csv}};
\addplot[cieee5!20!black,fill=cieee5!60!white] table[x index=0,y index=9,restrict expr to domain={\thisrow{show}}{1:1}] {\tables{power_breakdownAll.csv}};



  





\legend{Memory,	ChSum,	SoPs,	ImgBank,	FilterBank,	Other} 

\end{axis} \end{tikzpicture}
\caption{Core power breakdown for fixed-point and several binary architectures.}
\label{fig:barpower}
\end{figure}

%% file: realapplication.tex
\begin{table*}[ht!]
	\caption{Evaluation on Several Widely-Known Convolutional Neural Networks in the High-Efficiency Corner}
	\label{tbl:realapp}
	\centering
	\footnotesize
	\begin{tabularx}{2.0\columnwidth}{l|rrrrrrr|rrrrrr|rr}
\toprule
\multirow{2}{*}{Network} & $L$ & $h_k$ & $w$ & $h$ & $n_{in}$ & $n_{out}$ & $\times$  & $\eta_{tile}$ & $\eta_{Idle}$ & $\tilde{P}_{real}$ & $\Theta_{real}$ & $EnEff$  & & $t$ & $E$ \\ 
 & & px & px & px & & & & &  &  & GOp/s & TOp/s/W & \#MOp & ms & \textmu J \\ \midrule
\multirow{6}{*}{\begin{minipage}{16mm}BinaryConnect\newline Cifar-10 \cite{Courbariaux2015a}\end{minipage}} & 1                 & 3 & 32  & 32  & 3   & 128 & 1    & 1.00 & 0.09 & 0.35 & 1.9  & 16.0 & 7    & 3.8   & 0.4  \\
 & 2                 & 3 & 32  & 32  & 128 & 128 & 1    & 1.00 & 1.00 & 1.00 & 20.1 & 59.2 & 302  & 15.0  & 5.1  \\
 & 3                 & 3 & 16  & 16  & 128 & 256 & 1    & 1.00 & 1.00 & 1.00 & 20.1 & 59.2 & 151  & 7.5   & 2.6  \\
 & 4                 & 3 & 16  & 16  & 256 & 256 & 1    & 1.00 & 1.00 & 1.00 & 20.1 & 59.2 & 302  & 15.0  & 5.1  \\
 & 5                 & 3 & 8   & 8   & 256 & 512 & 1    & 1.00 & 1.00 & 1.00 & 20.1 & 59.2 & 151  & 7.5   & 2.6  \\
 & 6                 & 3 & 8   & 8   & 512 & 512 & 1    & 1.00 & 1.00 & 1.00 & 20.1 & 59.2 & 302  & 15  & 5.1  \\
 & 7                 & FC & 4 & 4 & 512 & 1024 & 1 & & & & & & 16 & & \\
 & 8                 & FC & 1 & 1 & 1024 & 1024 & 1 & & & & & & 2 & & \\
 & 9                 & SVM & 1 & 1 & 1024 & 10 & 1 & & & & & & 0.0 & &\\
\midrule
\multirow{3}{*}{\begin{minipage}{16mm}BinaryConnect\newline SVHN \cite{Courbariaux2015a}\end{minipage}} & 1                 & 3 & 32  & 32  & 3   & 128 & 1    & 1.00 & 0.09 & 0.35 & 1.9  & 16.0 & 7    & 3.8   & 0.4  \\
 & 2                 & 3 & 16  & 16  & 128 & 256 & 1    & 1.00 & 1.00 & 1.00 & 20.1 & 59.2 & 151  & 7.5   & 2.6  \\
 & 3                 & 3 & 8   & 8   & 256 & 512 & 1    & 1.00 & 1.00 & 1.00 & 20.1 & 59.2 & 151  & 7.5   & 2.6  \\
 & 4                 & FC & 4 & 4 & 512 & 1024 & 1 & & & & & & 16 & & \\
\midrule
\multirow{5}{*}{\begin{minipage}{16mm}AlexNet\newline ImageNet\cite{Krizhevsky2012}\end{minipage}} & 1ab\footnotemark & 6 & 224 & 224 & 3 & 48 & 4 & 0.95 & 0.09 & 0.35 & 1.4 & 12.1 & 520 & 364.7 & 42.9 \\
 & 1cd\textsuperscript{\ref{footnotehk11}} & 4 & 224 & 224 & 3 & 48 & 4 & 0.9 & 0.07 & 0.35 & 3.55 & 11.8 & 361 & 101.7 & 30.5 \\
 & 2                 & 5 & 55  & 55  & 48  & 128 & 2    & 0.93 & 0.75 & 1.00 & 39.1 & 45.2 & 929  & 23.8  & 20.6 \\
 & 3                 & 3 & 27  & 27  & 128 & 192 & 2    & 1.00 & 1.00 & 1.00 & 20.1 & 59.2 & 322  & 16.0  & 5.4  \\
 & 4                 & 3 & 13  & 13  & 192 & 192 & 2    & 1.00 & 1.00 & 1.00 & 20.1 & 59.2 & 112  & 5.6   & 1.9  \\
 & 5                 & 3 & 13  & 13  & 192 & 128 & 2    & 1.00 & 1.00 & 1.00 & 20.1 & 59.2 & 75   & 3.7   & 1.3  \\
 & 7                 & FC & 13 & 13 & 256 & 4096 & 1 & & & & & & 354 & & \\
 & 8                 & FC & 1 & 1 & 4096 & 4096 & 1 & & & & & & 34 & & \\
  & 9                 & FC & 1 & 1 & 4096 & 1000 & 1 & & & & & & 8 & & \\
\midrule                     
\multirow{8}{*}{\begin{minipage}{16mm}ResNet-18/34\newline ImageNet\cite{He2015}\end{minipage}} & 1                 & 7 & 224 & 224 & 3   & 64  & 1    & 0.86 & 0.09 & 0.35 & 4.4  & 15.1 & 236  & 53.3 & 15.7 \\
 & 2-5               & 3 & 56 & 56 & 64  & 64  & 4/6  & 0.95 & 1.00 & 1.00 & 19.1 & 56.2 & 231  & 11.9  &  4.0 \\
 & 6                 & 3 & 28  & 28  & 64  & 128 & 1    & 0.97 & 1.00 & 1.00 & 19.4 & 57.2 & 116  &  5.7  & 2.0  \\
 & 7-9              & 3 & 28  & 28  & 128 & 128 & 3/7  & 0.97 & 1.00 & 1.00 & 19.4 & 57.2 & 231  & 11.5  &  3.9 \\
 & 10                & 3 & 14  & 14  & 128 & 256 & 1    & 1.00 & 1.00 & 1.00 & 20.1 & 59.2 & 116  &  5.7  & 2.0 \\
 & 11-13             & 3 & 14  & 14  & 256 & 256 & 3/11 & 1.00 & 1.00 & 1.00 & 20.1 & 59.2 & 231  & 11.5  &  3.9 \\
 & 14                & 3 & 7  & 7  & 256 & 512 & 1    & 1.00 & 1.00 & 1.00 & 20.1 & 59.2 & 116  &  5.7  & 2.0  \\
 & 15-17             & 3 & 7  & 7  & 512 & 512 & 3    & 1.00 & 1.00 & 1.00 & 20.1 & 59.2 & 231  & 11.5  &  3.9 \\
  & 18                 & FC & 7 & 7 & 512 & 1000 & 1 & & & & & & 200 & & \\
\midrule     
\multirow{9}{*}{\begin{minipage}{16mm}VGG-13/19\newline ImageNet\cite{Simonyan2014}\end{minipage}}  & 1                 & 3 & 224 & 224 & 3   & 64  & 1    & 0.95 & 0.09 & 0.35 & 1.9  & 15.2 & 173  & 91.9  & 11.4 \\
 & 2                 & 3 & 224 & 224 & 64  & 64  & 1    & 0.95 & 1.00 & 1.00 & 19.1 & 56.2 & 3699 & 193.6 & 65.8 \\
 & 3                 & 3 & 112 & 112 & 64  & 128 & 1    & 0.95 & 1.00 & 1.00 & 19.1 & 56.2 & 1850 & 96.8  & 32.9 \\
 & 4                 & 3 & 112 & 112 & 128 & 128 & 1    & 0.95 & 1.00 & 1.00 & 19.1 & 56.2 & 3699 & 193.6 & 65.8 \\
 & 5                 & 3 & 56  & 56  & 128 & 256 & 1    & 0.97 & 1.00 & 1.00 & 19.4 & 57.2 & 1850 & 95.2  & 32.4 \\
 & 6                 & 3 & 56  & 56  & 256 & 256 & 1/3  & 0.97 & 1.00 & 1.00 & 19.4 & 57.2 & 3699 & 190.3 & 64.7 \\
 & 7                 & 3 & 28  & 28  & 256 & 512 & 1    & 1.00 & 1.00 & 1.00 & 20.1 & 59.2 & 1850 & 91.9  & 31.2 \\
 & 8                 & 3 & 28  & 28  & 512 & 512 & 1/3  & 1.00 & 1.00 & 1.00 & 20.1 & 59.2 & 3699 & 183.8 & 62.5 \\
 & 9-10              & 3 & 14  & 14  & 512 & 512 & 2/4  & 1.00 & 1.00 & 1.00 & 20.1 & 59.2 & 925  & 45.9  & 15.6\\
  & 11                 & FC & 14 & 14 & 256 & 4096 & 1 & & & & & & 411 & & \\
 & 12                 & FC & 1 & 1 & 4096 & 4096 & 1 & & & & & & 34 & & \\
  & 13                 & FC & 1 & 1 & 4096 & 1000 & 1 & & & & & & 8 & & \\
\bottomrule
\end{tabularx}\\
\vspace{2mm}
\textbf{Legend: } L: layer, $h_k$: kernel size, w: image width, h: image height, $n_{i}$: input channels, $n_{o}$: output channels, $\times$: quantity of this kind of layer, \newline $\eta_{tile}$: tiling efficiency, $\eta_{chIdle}$: channel idling efficiency, $\tilde{P}_{real} $ Normalized Power consumption in respect to active convolving mode, \newline $\Theta_{real}$: actual throughput, EnEff: Actual Energy Efficiency, \#MOp: Number of  operations (additions or multiplications, in millions), t: time, E: needed processing energy
\end{table*}

%% file: soa.tex
\begin{figure}
\centering
\begin{tikzpicture}[scale=0.75]
\tikzset{mark options={solid, mark size=3, line width=1pt}} %



	\begin{loglogaxis}[width=1.33\columnwidth-1.5cm, height=7cm, 
	    log basis x=10, 
		xlabel={Core Area Efficiency {[GOp/s/MGE]}},ylabel={Core Energy Efficiency {[GOp/s/W]}},  grid=major,  ymax=100000, ymin=100, xmin=1]

\addplot[color=blue,mark=*,only marks] table [x=areaeff, y=eneff, col sep=space, unbounded coords=discard, restrict expr to domain={\thisrow{show}}{3:3}]{\tables{soa_plot.csv}}; \label{soaAreaEnergy:ASIC}
 \addplot[color=red,only marks] table [x=areaeff, y=eneff, col sep=space, unbounded coords=discard, restrict expr to domain={\thisrow{show}}{5.1:5.1}]{\tables{soa_plot.csv}}; \label{soaAreaEnergy:Yoda12}
 \addplot[color=red,mark=ball, ball color=red!80!black, only marks] table [x=areaeff, y=eneff, col sep=space, unbounded coords=discard, restrict expr to domain={\thisrow{show}}{5.2:5.2}]{\tables{soa_plot.csv}}; \label{soaAreaEnergy:Yoda06}
    
   \addplot[color=black, only marks, sharp plot, style=dashed] table [x=areaeff, y=eneff, col sep=space, unbounded coords=discard, restrict expr to domain={\thisrow{show}}{6:6}]{\tables{soa_plot.csv}}; \label{soaAreaEnergy:pareto}  
   
   \addplot[color=black, only marks, sharp plot, style=densely dotted] table [x=areaeff, y=eneff, col sep=space, unbounded coords=discard, restrict expr to domain={\thisrow{show}}{7:7}]{\tables{soa_plot.csv}}; \label{soaAreaEnergy:paretoPrev}  
   
   \node at (axis cs:41.4,5000) [anchor=south west] {\protect\NoHyper{EIE \cite{Han2016b}}};
   \node at (axis cs:109.6,1930) [anchor=south west] {\protect\NoHyper{k-brain \cite{kbrain2015}}};
   
   \node at (axis cs:48,1800) [anchor=east] {\protect\NoHyper{\cite{park2016}}}; 

\node at (axis cs:1.4,465) [anchor=south west] {\protect\NoHyper{\cite{Reagen2016}}}; 
\node at (axis cs:14.4,960) [anchor=east] {\protect\NoHyper{\cite{Likamwa2016}}}; 
\node at (axis cs:20,1420) [anchor=south east] {\protect\NoHyper{\cite{Jaehyeong2016}}}; 
\node at (axis cs:34.3,800) [anchor=east] {\protect\NoHyper{\cite{Cavigelli2016}}}; 
\node at (axis cs:90.7,437) [anchor=north east] {\protect\NoHyper{\cite{Cavigelli2016}}}; 
\node at (axis cs:42,630) [anchor=north east] {\protect\NoHyper{\cite{Du2015}}}; 

\node at (axis cs:16.6,490) [anchor=north east] {\protect\NoHyper{\cite{Pham:2012_Neuflow}}}; 

\node at (axis cs:2.9,380) [anchor=south west] {\protect\NoHyper{\cite{Shafiee2016}}}; 
   
    	\end{loglogaxis}



  


\end{tikzpicture}\\
\begin{scriptsize}	
		SoA ASICs (\ref{soaAreaEnergy:ASIC}), 
		YodaNN (1.2--0.6V) (\ref{soaAreaEnergy:Yoda12}), 
		YodaNN (1.2 \& 0.6V corner) (\ref{soaAreaEnergy:Yoda06}), \\
		previous Pareto front (\ref{soaAreaEnergy:paretoPrev}), 
		our Pareto front (\ref{soaAreaEnergy:pareto})
	\end{scriptsize}

\caption{\revAdd{Core area efficiency vs. core energy efficiency for state-of-the-art CNN accelerators.}}\label{fig:soaAreaEnergy}
\end{figure}
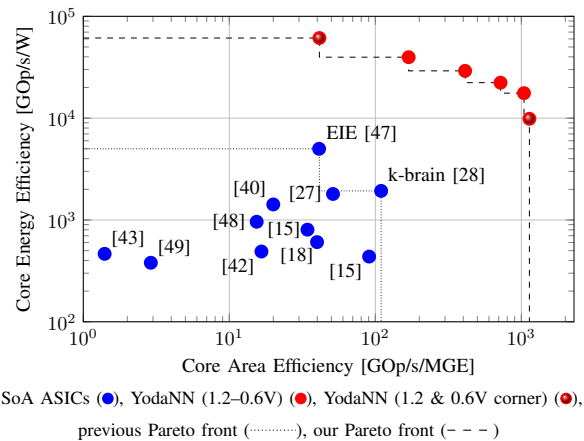

%% file: 5_conclusion.tex
\section{Conclusion}
We have presented a flexible, energy-efficient and performance scalable CNN accelerator. The proposed architecture is the first ASIC design exploiting recent results on \new{binary-weight} CNNs, which greatly simplifies the complexity of the design by replacing fixed-point MAC units with simpler complement operations and multiplexers without negative impact on classification accuracy. To further improve energy efficiency and extend the performance scalability of the accelerator, we have implemented latch-based SCMs for on-chip data storage to be able to scale down the operating voltage even further. To add flexibility, we support seven different kernel sizes: \revAdd{$1\times 1$, $2\times 2$, ..., $7\times 7$.} This enables efficient evaluation of a large variety of CNNs. Even though this added flexibility introduces a \calc{0}{(1-108/153)*100}\% reduction in energy efficiency, an outstanding overall energy efficiency of \SI{61}{TOp/s/W} is achieved. \revAddC{The proposed accelerator surpasses state-of-the-art CNN accelerators by 2.7$\times$ in peak performance with \SI{1.5}{\tera Op/\second}, by 10$\times$ in peak area efficiency with \SI{1.1}{TOp/s/MGE} and by 32$\times$ peak energy efficiency with \SI{61.2}{\tera Op/\second/\watt}.} YodaNN's power consumption at \SI{0.6}{\volt} is \SI{895}{\micro\watt} with an average frame rate of \SI{11}{FPS} for state-of-the-art CNNs and \SI{16.8}{FPS} for ResNet-34 at \SI{1.2}{\volt}.


%% file: 6_futurework.tex

%% file: 7_acknowledgment.tex
\section*{Acknowledgments}
This work was funded by the Swiss National Science Foundation under grant 162524 (MicroLearn: Micropower Deep Learning), armasuisse Science \& Technology and the ERC MultiTherman project (ERC-AdG-291125). 
%
%
